\documentclass[preprintnumbers,reprint,twocolumn,superscriptaddress, amsmath,amssymb,aps,nofootinbib]{revtex4-1}
\usepackage{graphicx}
\usepackage{dcolumn}
\usepackage{bm}
\usepackage{float}
\usepackage{hyperref}
\usepackage{dsfont}
\usepackage{slashed}
\usepackage{color}
\usepackage{amsmath}
\usepackage[section]{placeins}
\usepackage{braket}
\usepackage{upgreek}
\usepackage[bottom]{footmisc}
\usepackage[normalem]{ulem}

\newcommand{\beq}{\begin{eqnarray}}
\newcommand{\eeq}{\end{eqnarray}}
\newcommand{\beqnn}{\begin{eqnarray*}}
\newcommand{\eeqnn}{\end{eqnarray*}}

\newcommand{\tr}{\text{Tr}}

\def\bs{\boldsymbol}

\long\def\comment#1{ }

\def\be{\begin{eqnarray*}}
\def\ee{\end{eqnarray*}}
\def\beq{\begin{eqnarray}}
\def\eeq{\end{eqnarray}}

\def\rmd{{\rm d}}

\def\cs{{\rm cs}}

\def\out{{\rm loss}}
\def\med{{\rm med}}

\def\bfk{{\bs k}}
\def\bfq{{\bs q}}
\def\bfu{{\bs u}}
\def\bfr{{\bs r}}

\begin{document}

\title{An EFT approach to Color decoherence in jet quenching}

\author{Varun Vaidya}
\affiliation{Department of Physics, 
University of South Dakota, 
SD, 57069, USA}
\email{Varun.Vaidya@usd.edu}

\date{\today}


\begin{abstract}
We use the EFT developed in \cite{Mehtar-Tani:2025xxd}, to understand the interference driven phenomenon of color decoherence in inclusive jet production in a dense nuclear medium such as Nuclei or Quark Gluon Plasma.  Using the factorization formula in \cite{Mehtar-Tani:2024smp,Mehtar-Tani:2025xxd}, expressed as a series of multi-sub-jet operators, we define and calculate the contribution of the two sub-jet effective operator. This explicitly reveals the emergent angular scale $\theta_c$ that controls color decoherence as well an intricate renormalization group structure for the factorized functions. We show that for a jet of radius R in a medium of size L characterized by a jet quenching parameter $\hat q$,  both the LPM effect and color decoherence are controlled by a single dimensionless parameter $\sqrt{\hat q L}LR$ and therefore are equally important for phenomenology. This paper shows how interference driven emergent effects can be included in a factorized framework for computing jet observables in heavy ion collisions.

\begin{description}
\item[Keywords]
Heavy Ion Phenomenology, Jets, SCET
\end{description}
\end{abstract} 

\maketitle


\section{Introduction}
 The phenomenon of ``jet quenching'', in which high energy jets lose energy while traversing a Quark Gluon Plasma(QGP) created in Heavy Ion Collisions was proposed in \cite{Bjorken:1982tu} and later successfully observed in Au-Au collisions at RHIC and at the LHC~\cite{BRAHMS:2004adc,PHOBOS:2004zne,STAR:2005gfr,PHENIX:2004vcz,ATLAS:2010isq,
ALICE:2010yje,
CMS:2011iwn,ATLAS:2018gwx,CMS:2021vui,ALICE:2023waz,Connors:2017ptx}. There has been increasing interest in using a high energy jet as a hard probe of the medium properties of a QGP created in Heavy Ion collisions as well as Heavy Nuclei in Electron Ion collisions. However, the theory of jet quenching has yet to attain the predictive power of simpler systems, such as electron-positron or $pp$ collisions. 

 Quantitative precision in pp collisions~\cite{Currie:2016bfm,Czakon:2019tmo}has been achieved through factorization ~\cite{Collins:1989gx,Larkoski:2017jix} that separates perturbative dynamics at short distances from non-perturbative physics at long-distance scales through universal functions such as parton distribution functions (PDFs) for the initial state or shape functions~\cite{Korchemsky:1999kt,Lee:2006nr} for the final state. These factorization formulas can be derived using Effective Field Theory (EFT) tools~\cite{Larkoski:2017jix,Asquith:2018igt,Marzani:2019hun} such as Soft Collinear Effective Theory (SCET) \cite{Bauer:2002aj,Bauer:2003mga,Bauer:2000yr,Bauer:2001ct,Bauer:2002nz}. 

Factorization is especially relevant for heavy ion collisions and electron ion collisions where the jet propagates in a strongly coupled medium and an operator based encoding of non-perturbative physics is crucial for predictive power. However, factorization in jet quenching is far more challenging than in pp due to the multitude of new emergent phenomenon and the corresponding scales. This is a direct result of the interaction of the jet partons with an \textit{extended} medium which requires us to treat the jet as an Open quantum system.
For example, an important phenomenon in partonic energy loss is the Landau-Pomeranchuk-Migdal (LPM) effect \cite{Landau:1953um,Migdal:1956tc}, which  is controlled by an emergent scale known as the parton formation time. This was first understood in the 1990s~\cite{Gyulassy:1993hr,Wang:1994fx,Baier:1994bd,Baier:1996kr,Baier:1996sk,Zakharov:1996fv,Zakharov:1997uu,Gyulassy:2000er,Wiedemann:2000za,Guo:2000nz,Wang:2001ifa,Arnold:2002ja,Arnold:2002zm} and phenomenological models were developed to describe the data~\cite{Salgado:2003gb,Liu:2006ug,Qin:2007rn,Armesto:2011ht}. Similarly, because jets are extended multi-partonic systems, interference between multiple fast-moving color charges occurs in the plasma, which is controlled by the angular resolution power of the QGP known as the critical angle $\theta_c$. The corresponding phenomenon is called color (de)coherence has been explored in fixed order calculations in certain limits~\cite{Mehtar-Tani:2010ebp,Mehtar-Tani:2012mfa,Casalderrey-Solana:2011ule, Casalderrey-Solana:2012evi, Mehtar-Tani:2011hma,Mehtar-Tani:2017ypq} as well as resulting non linear evolution \cite{Mehtar-Tani:2024mvl}. Its importance for phenomenology was also explored \cite{Mehtar-Tani:2014yea,Mehtar-Tani:2021fud}. However, an EFT based factorization formula for a complete calculation of specific jet observables that systematically separates physics by scale and accounts for these interference-driven phenomenon is currently lacking.

This problem was recently addressed through an Open quantum system EFT framework proposed in \cite{Mehtar-Tani:2024smp,Mehtar-Tani:2025xxd}. The observable considered was inclusive jet production of a narrow jet $R\ll 1$, which is the foundation for future exploration of jet substructure observables. The relevant cross section is a histogram of jets based on their transverse momentum $p_T$ and rapidity $\eta$, for a fixed jet radius $R$. These works laid down the framework to understand how the physics at distinct scales, including any emergent scales, can be factorized in terms of gauge invariant operators. Most notably, it showed that capturing all the novel interference driven phenomenon in the medium requires an infinite series of multi-sub-jet operators. This was a significant extension of previous results in the literature where EFT techniques were applied to jets in HIC ~\cite{Idilbi:2008vm,DEramo:2010wup,Ovanesyan:2011xy}. The explicit calculations in \cite{Mehtar-Tani:2025xxd} were restricted to a dilute medium for a single sub-jet. 

In this paper, using the same EFT approach, we explore the landscape of multi sub-jet operators which is important for understanding several important effects: the contribution of non-global logarithms, the LPM effect for multi-partonic jets, emergence of color decoherence and finally the renormalization group evolution of the factorized functions. We explicitly compute the matching co-efficients,  the leading fixed order result for both vacuum and medium evolution as well as the leading renormalization group equation for the one and two sub-jet operators.  In this process, we show how the scale $\theta_c$ emerges in our EFT in the same way as in a full QCD calculation for a dipole \cite{Mehtar-Tani:2012mfa}. We also show that both the LPM effect and color decoherence for this observable depend on a single dimensionless parameter and that multi-sub-jet  operators contribute at the same order in power counting as the single sub-jet and must be included for precision phenomenology. 

This paper is organized as follows. We first describe the EFT framework in Section \ref{sec:EFT}. We then compute the one loop results as well the leading Renormalization Group(RG) equations for the matching co-efficients for the one and two sub-jet operators in Section \ref{sec:Match}. In Section \ref{sec:Oneloop}, we present fixed order results for the one and two sub-jet operators in vacuum and for single interaction with the medium, and discuss the emergence of $\theta_c$. Consistency constraints on RG evolution are described in Section\ref{sec:cfact} followed by a summary of our results and discussion of open questions in Section \ref{sec:Disc}.

\section{EFT for inclusive jet production}
\label{sec:EFT}
In \cite{Mehtar-Tani:2024smp}, we developed an EFT for describing jet evolution in heavy ion collisions. The final state observable was a jet with radius $R \ll 1$. 
We identified three relevant well separated scales in this system $p_T \gg p_TR \gg Q_{\text{med}} \geq m_D$ and wrote a factorization formula decoupling the physics at these scales. Here $Q_{\text{med}}$ is the average transverse momentum exchange by the medium with a high energy parton. For a medium of size L, this is parameterized in literature in terms of the jet quenching parameter $\hat q$ as $Q^2_{\text{med}} = \hat q L$. 
Phenomenologically, we can understand the evolution of the jet in terms of these scales as follows. 
The parton initiating the jet is a high energy particle created through a large momentum transfer ($p_T \sim 100$s GeV). 
This parton evolves by splitting into multiple partons with an energy fraction O(1) at an angle dictated by the measurement, in this case the jet radius $R \ll 1$. The phase space for these type of splittings is therefore characterized by hard collinear partons whose momentum in the light-cone co-ordinates scales as $p_{hc} \sim p_T(1, R^2,R)$. The corresponding momentum transfer during this splitting ($p_TR \sim 10$s GeV) is still much higher than the typical momentum exchange with the medium and hence this stage of the evolution is unaffected by the medium and happens in the vacuum.
The transverse momentum that the medium typically transfers to a high energy parton over multiple interactions with it is given by $Q_{\text{med}} \sim 1$(s) GeV  which is a function of the temperature T of the medium and its density. We note that any such interaction of the medium with a jet parton is again constrained by the measurement to preferentially split the parton at an angle R. This dictates that the energy of the split parton must scale as $\sim Q_{\text{med}}/R \sim 10$(s) GeV, which are softer than the hard-collinear partons. We call the phase-space for these emissions as collinear soft $p_{cs} \sim Q_{\text{med}}/R(1, R^2, R)$. These collinear soft emissions that are sourced either through vacuum or medium induced evolution can further re-scatter off the medium partons which are soft and have momenta that scale as $p_s \sim T(1, 1, 1)$. In a strongly coupled medium $T \sim m_D \leq Q_{\text{med}}$. Therefore the physics at each scale or equivalently, virtuality, is dictated by specific regions in momentum phase space.
 
Based on this separation, a factorization formula at leading power in the expansion parameter R, was written in terms of a single hard function $H_i$ which describes, at the scale $\omega_J \sim p_T$, the production of a high energy parton of species i that initiates the jet.  This is convolved with a jet function $J_i$ that describes the evolution of the jet at scales $p_TR $ and below. 
\begin{equation}
\!\!\! \frac{{\rm d}\sigma}{{\rm d}p_T{\rm d}\eta}= \sum_{i \in q, \bar q , g}\int_0^1 \frac{{\rm d}z}{z} \, H_i\left(\frac{\omega_J}{z},\mu\right) \, J_i(z, \omega_J, R,\mu) +{\cal O}(R^2)
\label{eq:factI}
\end{equation}
Here z is the energy fraction of the parton created in the hard scattering that remains inside the jet while $\omega_J$ is the jet energy. The momentum fraction $z$ is related to the jet $p_T$ and rapidity $\eta$ via $\omega_J = 2p_T \cosh \eta$. 

\begin{figure}
\centering
\includegraphics[width=0.6\linewidth]{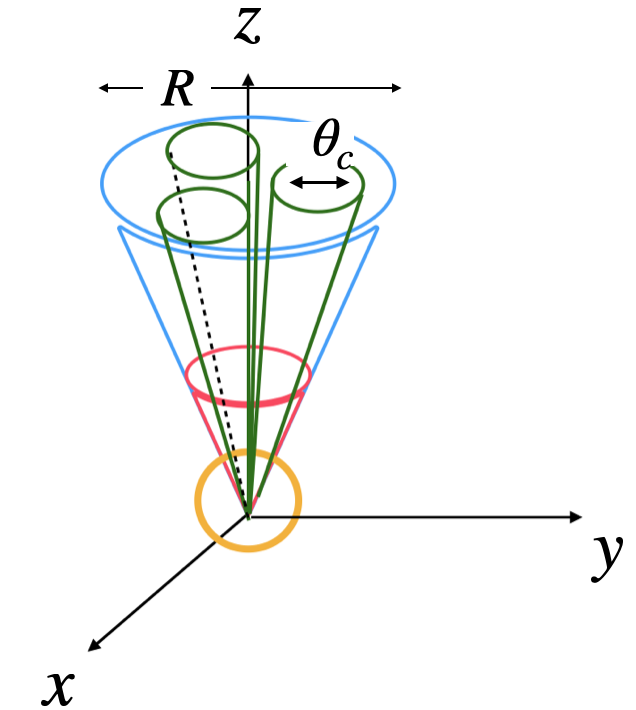}
\caption{Phase space structure for jet evolution in the medium. The medium resolves multiple hard collinear partons(green) inside the jet (blue). Interaction of these subjets with medium partons(orange) leads to collinear soft radiation(red).}
\label{fig:sj}
\end{figure}

As the high energy jet partons move through the medium, they interact by exchanging gluons with the medium partons. This puts them off-shell by $p^2 \leq Q_{\text{med}}^2$ allowing them to radiate collinear-soft gluons. One would then expect that in a multi-partonic jet, every single hard collinear parton would act as an independent color source of medium induced radiation and therefore the energy loss should grow with increasing number of jet partons. However, we also recognize that if the partons are not well separated, we would expect strong quantum interference leading to suppression of medium induced radiation. This is dictated by the ability of the medium to resolve partonic color through an emergent scale known as the critical angle $\theta_c$. This can be thought of as the smallest angular separation between partons that can be resolved by the medium. Since the medium exchanges a transverse momentum $\sim Q_{\text{med}}$ with the jet partons, it can resolve a transverse distance of $\mathcal{O}(1/Q_{\text{med}})$. The measurement imposed on the jet, namely its jet radius R dictates the typical angular separation between the jet partons. Thus their transverse distance over the medium length grows to $LR$. Therefore such partons can be resolved if $R L \gg 1/Q_{\text{med}}$, which defines the critical angle as $\theta_c = 1/(Q_{\text{med}}L)$ as shown in Fig.~\ref{fig:sj}.  Thus color decoherence in controlled by the dimensionless parameter $ \lambda_m = R/\theta_c \sim RQ_{\med}L$.  In a large and dense medium which one may expect in Pb-Pb or Au-Au collisions, phenomenological studies tell us that $Q_{\text{med}}\sim 1- 2$ GeV in a medium of size $\sim 5$ fm. This corresponds to a $\theta_c \sim 0.01 \sim 0.02 \ll R$. Therefore we expect the medium to resolve multiple sub-jets. 

There is another scale in the problem associated with purely the vacuum evolution, which is $Q_T\sim (1-z) \omega_JR$ and in order to write down the Stage II factorization, we need to understand how this compares with the medium scale $Q_{\text{med}}$. For the regime when $z\sim 1$ but is not very close to 1, $Q_T \sim p_TR$ and therefore vacuum evolution stops at this scale. On the other hand when $z \rightarrow 1$, which is the threshold limit, we can have $Q_T \sim Q_{\text{med}}$, which then requires us to also separate vacuum evolution at the scale $\omega_JR $ from $Q_T$. From Eq.~\ref{eq:factI}, we see that the integral over z covers both these regimes so we need to consider both types of factorizations and smoothly interpolate between them. In this paper, we analyze the region $z \rightarrow 1$, leaving the $z \sim 1$ regime for future work. Therefore we now write down a factorization at leading power in both R and $Q_{\text{med}}/(\omega_JR) \sim (1-z)$.

To correctly include these effects, we need to modify the Hard function $H \rightarrow \hat H$ keeping only leading order terms in $z \rightarrow 1$. At the same time, we need to introduce a global soft mode that scales as $p_{GS} \sim p_TR(1,1,1)$ and recoiling collinear function $J_X$ which also contributes to jet energy loss. The factorization now takes the form
\begin{widetext}
\begin{eqnarray}
\!\!\! \frac{{\rm d}\sigma}{{\rm d}p_T{\rm d}\eta}&=& \sum_{i \in q, \bar q , g}\int_0^1 \frac{{\rm d}z}{z}\int \rmd \epsilon_S\ \, \langle \hat H_i\left(\frac{\omega_J}{z},\mu\right) \, S_G(\epsilon_S,\mu) \rangle\int_{0}^{1} \rmd z' \,\int\rmd \epsilon \, \delta(\omega_J'-\omega_J-\epsilon- \epsilon_S-\epsilon_X) J_X(\epsilon_X,\mu) \nonumber\\
& &\sum_{m =1}^{\infty} \prod_{j=2}^m\int \frac{{\rm d}\Omega(n_j)}{4\pi} \langle\mathcal{C}_{i\rightarrow m}\Big(\{\underline{n}\},z', \omega_J'= \frac{z'\omega_J}{z}, \mu\Big)\, {\cal S}_{m} (\{\underline{n}\} , \epsilon,\mu)\rangle + {\cal O}(R^2)+O\left(\frac{Q^2_{\text{med}}}{(p_TR)^2} \sim (1-z)^2\right)\nonumber\\
& \equiv & \langle H\times S_G \rangle \otimes_z J_X \otimes_z \sum_{m=1}^{\infty} Q_i
\label{eq:factI2}
\end{eqnarray}
\end{widetext}
which holds up to power corrections ${\cal O}(Q_{\rm med}/(p_TR))\sim O(1-z)$. Here, $\mu$ is the renormalization scale. The color indices of the cs operator ${\cal S}_m$ are contracted with the corresponding Wilson co-efficient ${\cal C}_{i \rightarrow m}$ which is denoted by the  color trace $\langle .. \rangle$. In the same way color indices of the global soft function $S_G$ are contracted with the Hard function. We can interpret ${\cal C}_{i \rightarrow m}$ as the probability to create m final state hard collinear partons mutually separated by $\theta \sim R$ inside the jet. Note that since we are in the threshold limit, ${\cal C}_{i \rightarrow m}(z')$ is always proportional to $\delta(1-z')$ so that ultimately it does not contribute to energy loss. For an m sub-jet configuration, we can identify m light-like vectors $\{n_1,n_2, .. ,n_m\}$ along the directions of hard partons that source these sub-jets. 
We note that the form of this factorization is the same as proposed in \cite{Liu:2017pbb} for the joint R and threshold resummation of semi-inclusive jets. This accounts for any non-global logarithmic contributions to the the vacuum evolution. The definitions of the functions  $\hat H_i, S_G, {\cal C}_{i\rightarrow m} ,J_X$, which do not involve medium interactions, are already known in literature\cite{Liu:2017pbb} from their computation in pp collisions and remain unchanged. On the other hand the collinear soft function which involves medium-jet interactions is modified. We define our operator ${\cal S}_m$ as,
\begin{widetext}
\begin{align}
    \label{eq:soft-funct}
&{\cal S}_{m}^{i_0i'_0i_1i'_1i_2i'_2...i_mi'_m}(\{\underline{n}\},\epsilon,\mu) \equiv    \text{Tr}\Big[{\bf U}_{i_mj_m}^{(r_m)}(n_m)...{\bf U}_{i_1j_1}^{(r_1)}(n_1)U_{i_0j_0}(\bar n)\rho_M U^{\dag}_{i'_0j_0}(\bar n){\bf U}^{\dag, r_1}_{i'_1j_1}(n_1)...{\bf U}^{\dag, r_m}_{i'_mj_m}(n_m)\mathcal{M}\Big] \,,
\end{align}
\end{widetext}
where $\{\underline{n}\}\equiv\{n_1,n_2,...,n_m\} $ denotes the directions of the $m$ collinear subjets. $n= (1,0,0,1) $ denotes the direction of the jet axis with the conjugate direction $\bar n(1,0,0,-1)$. We note that ${\cal S}_m$ is a matrix in color space. $\rho_M$ is the density matrix of the medium.

The measurement is defined as $\mathcal{M}= \Theta_{\text{alg}}\delta(\epsilon - \bar n \cdot p_\out )$ where $\epsilon$ is the energy loss due to radiation escaping the jet.
${\bf U}(n_i) = U_{\bf M}(n_i)U(n_i) $
where $U(n_i)$ is a path ordered Wilson line which reads
\beq 
U(n)\equiv {\bf P} \exp\left[ig \int_0^{+\infty} \rmd s  \, n\cdot A_{\rm cs}(s n)\right]\,.
\label{eq:CSWilson}
\eeq
The Wilson line $U_0(\bar n)$, which ensures gauge invariance, describes an unresolved effective charge moving in the opposite direction. The operator $U_{\bf M}^{r}(n)$ encodes medium induced collinear-soft radiation from a quark or gluon 
\beq 
U_{\bf M}^{r}(n)\equiv {\bf P} \exp\left[ \int_0^{+\infty} \rmd s  \, \int {\rm d}^2\bfq \frac{1}{\bfq^2}\Big[O^{ab}_{\text{cs}}\frac{1}{\mathcal{P}_{\perp}^2}\mathcal{O}_{\text{s}}^b\Big](sn,\bfq) t_r^a\,,\right] \nonumber 
\label{eq:CSMed}
\eeq 
The collinear soft function in Eq.\ref{eq:soft-funct} evolves with an effective Hamiltonian 
\begin{equation}
\label{eq:H2}
\int\! {\rm d}t\, \tilde H(t)\!=\!\int {\rm d}t \left(H_{\rm cs}(t)+H_{\rm s}(t)+ H_{\rm cs\text{-}s}(t)\right)
\end{equation}
 $H_{\text{cs}}$ is standard collinear SCET Hamiltonian, and $H_{\text{s}}$ describes the dynamics of the soft field. Moreover, $H_{\text{cs-s}}$ describes the interaction of the collinear-soft gluon with the soft medium parton, which we assume is dominated by forward scattering mediated by Glauber gluons.
The detailed definitions of these operators is given in the Appendix \ref{app:Ocs}. In the next section, we will compute the matching co-efficient ${\cal C}_{i\rightarrow m}$ for one and two sub-jet operators. 

\section{Matching coefficients for collinear soft operators}
\label{sec:Match}
\begin{figure}
\centering
\includegraphics[width=0.9\linewidth]{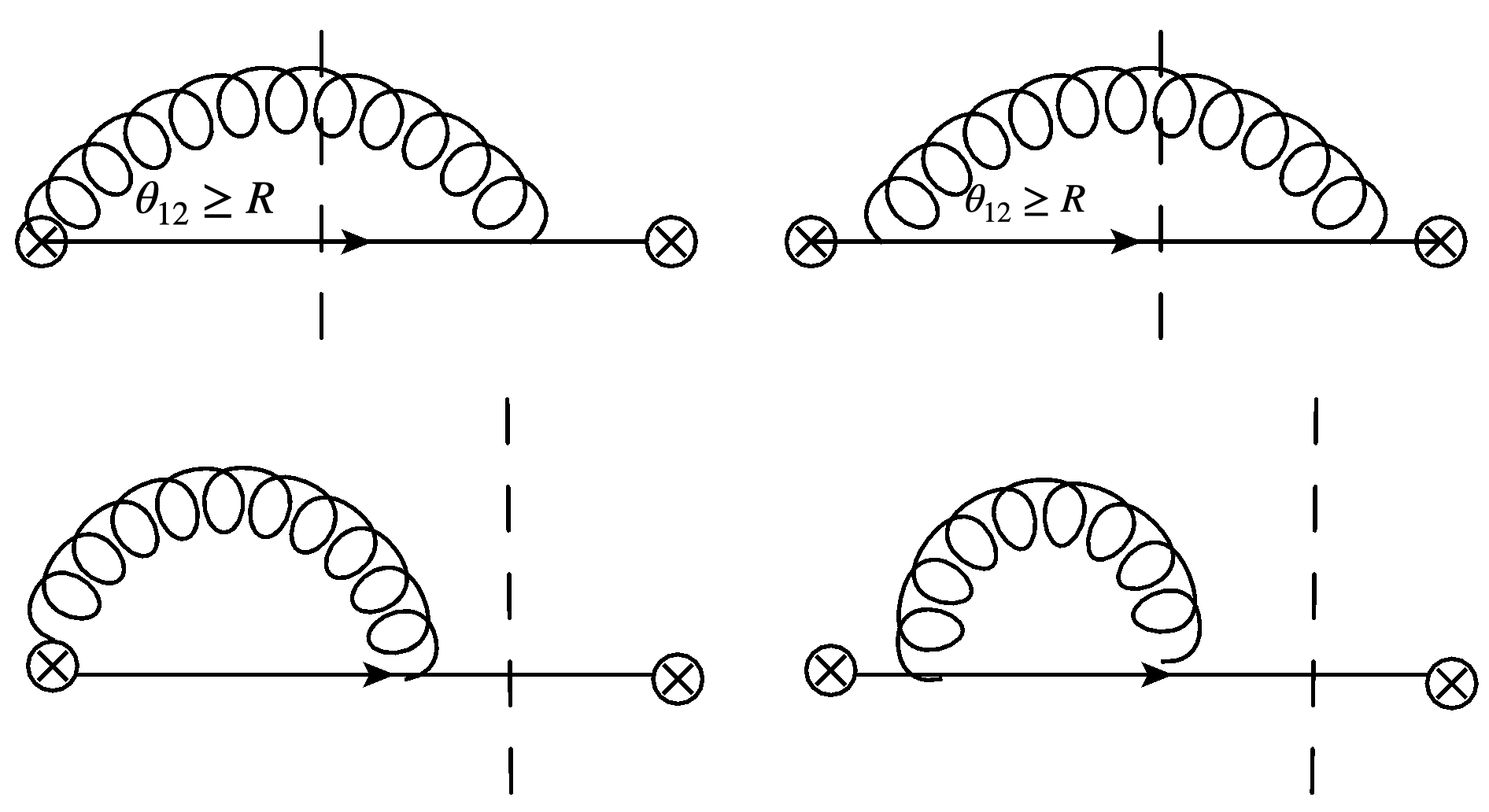}
\caption{Real and virtual diagrams of the collinear jet function that contribute to the single sub-jet quark matching co-efficient.}
\label{fig:C1}
\end{figure}

The matching co-efficients ${\cal C}_{i\rightarrow m}$ only encode vacuum evolution at the scale $p_TR$ and hence are identical to those for threshold jet production in pp collisions. A detailed computation for the thrust axis for a different observable was presented in \cite{Becher:2015hka} along with a Leading Log computation in \cite{Becher:2023znt} for the anti-kT algorithm. We present here the results for inclusive jet production at threshold. It was noted that these co-efficients are IR divergent and these divergences only cancel when we include  contributions from higher multiplicity sub-jet operators. In this paper we will re-organize the infinite series of sub-jet operators by redefining the collinear soft operators ${\cal S}_m$ so that the ${\cal C}_{i \rightarrow m}$ are IR finite and carry only UV divergences. \\
{\bf Matching for single sub-jet:} We first compute the matching co-efficient ${\cal C}_{q\rightarrow 1}$ for the single sub-jet operator ${\cal S}_1$. This corresponds to the first term in the series in Eq.\ref{eq:factI2}. 
\begin{eqnarray}
Q_1 &=& \sum_{r=q,g} {\cal C}^{j_0j'_0ii'}_{q \rightarrow r}(z', \omega'_J,\mu)\nonumber\\
&& \text{Tr}\Big[{\bf U}_{ji}^{(r)}(n)U_{j_0i_0}(\bar n)\rho_M U^{\dag}_{i_0j'_0}(\bar n){\bf U}^{(r)\dag}_{i'j}(n)\mathcal{M}\Big]
\end{eqnarray}
We will only consider the case of quark initiated jet in this paper for simplicity. The case for a gluon created in the hard interaction can be treated in the same way. 
In general therefore, $Q_1$ can be a sum over two terms; single quark or a single gluon inside the jet with the corresponding fundamental or adjoint representation for the ${\bf U}$ operator. We will refer to these two collinear soft operators as ${\cal S}_{1,q}$ and ${\cal S}_{1,g}$. 
For the quark operator, the color structure for the Wilson coefficient is simple so that we can write $
  {\cal C}^{j_0j'_0ii'}_{q\rightarrow q} = \delta^{ij_0}\delta^{i'j'_0}{\cal C}_{q\rightarrow q} $ where ${\cal C}_{q\rightarrow q}$ can be interpreted as the probability to create exactly one hard collinear quark inside the jet. Likewise ${\cal C}^{j_0j'_0ii'}_{q\rightarrow g} =[ t^i t^{i'}]_{j_0 j'_0}{\cal C}_{q\rightarrow g} $ where  $\mathcal{C}_{q\rightarrow g}$ is the probability for a single hard collinear gluon inside the jet.  Since we are considering a quark initiated jet, ${\cal C}_{q\rightarrow g} $ starts at O($\alpha_s$). For the threshold limit however, this corresponds to a collinear soft quark emission outside the jet and this is power suppressed so will vanish at leading power in the EFT. 
   Each collinear soft operator can be further expanded our order by order in the number of interactions with the medium. This is equivalent to the opacity expansion. For instance, we can write
\begin{eqnarray}
    \delta^{ij_0}\delta^{i'j^{'}_0}{\cal S}_{1,q}^{ii'j_0j_0'} = \sum_{k=0}^{\infty}{\cal S}_{q,k}
    \label{eq:SExp}
\end{eqnarray}
${\cal S}_{q,0}$ corresponds to vacuum evolution and is given by 
\begin{eqnarray}
 {\cal S}_{q,0}=  \text{Tr}\Big[ U(n)U(\bar n)\rho_M U^{\dag}(\bar n) U^{\dag}(n)\mathcal{M}\Big]
 \label{eq:Sq0}
\end{eqnarray} while ${\cal S}_{q,n}$ for $n \geq 1$ corresponds to n interactions with the medium. We will discuss the details of these terms in the next section.
  To compute the matching coefficients, as noted in \cite{Liu:2017pbb},  it is sufficient to match the vacuum evolution in the threshold limit ($z \rightarrow 1$) of the semi-inclusive jet function $J_i$ given in Eq.~\ref{eq:factI}. We can set $H_{\text{cs-s}}$, namely then interaction between the jet and the medium to zero to match the vacuum evolution. This requires us to compute ${\cal S}_{q,0}$ to one loop. The one loop result for ${\cal S}_{q,0}$ can be found in  \cite{Mehtar-Tani:2024smp}. 
  \begin{eqnarray}
\mathcal{S}^{(1)}_{q,0}(\epsilon_L)&=& \frac{\alpha_s\,C_F}{2\pi \omega_J'}\delta(1- \tilde z)\Big[-\frac{l^2}{2}+\frac{\pi^2}{12}\Big] \nonumber \\
&+&\frac{\alpha_s\,C_F}{2\pi \omega_J'}\Big[l\frac{2}{(1-\tilde z)_+}-4\left(\frac{\ln(1- \tilde z)}{1-\tilde z}\right)_+\Big], 
\label{eq:Sq0oneL}
\end{eqnarray}
where we have written $\epsilon_L= (1-\tilde z) \omega_J'$ and  $l$ is $\ln (4\mu^2/(\omega_JR)^2 )$. In Mellin space, ${\cal S}_{q,0}$ is function of $\ln (4\mu^2N^2/(\omega_JR)^2 ) $, where N is the Mellin variable conjugate to $(1-z)$. Therefore the natural scale for this function is $(1-z)\omega_JR/2 \ll \omega_JR/2$. 
  The computation of the jet function in the threshold limit involves four diagrams as shown in Fig.\ref{fig:C1}.
However, since we need exactly one quark inside the jet, we do not include the contribution when both the quark and gluon are inside the jet. This leads to a non-cancellation of IR divergence in the virtual diagram. While UV divergences are regulated using dim.reg. we do not regulate the IR divergences. To $\mathcal{O}(\alpha_s)$, this matching then yields 
  \begin{eqnarray}
&&\mathcal{C}_{q\to q}(z',\omega_J',\mu) =  \omega_J'\Bigg[\delta(1-z') \nonumber\\
&+& \delta(1-z')\frac{\alpha_s}{2\pi}C_F\Bigg[\frac{l^2}{2}+\frac{3l}{2}+\frac{13}{2}-\frac{3\pi^2}{4}\Bigg]\nonumber\\
&+& \omega'_J\delta(1-z')\frac{\alpha_sC_F}{\pi}\left(\frac{3}{2}+l \right)\int_0^R \frac{d\theta_{12}}{\theta_{12}}
\label{eq:CqIR}
\end{eqnarray} 
We see from the last line that this Wilson co-efficient has a collinear IR singularity as $\theta_{12} \rightarrow 0$ which is contributed by the virtual diagram. We will see in the next section how this IR singularity can be cured by a redefinition of the operators. 
Similarly as explained earlier, for the quark initiated jet, in the threshold limit, 
\begin{eqnarray}
&& {\cal C}_{q\rightarrow g}(z' ,\omega'_J, \mu) = 0
\end{eqnarray}
The natural scale for the matching co-efficient is $\mu \sim \omega_JR/2$, which is a perturbative scale. 

\begin{figure}
\centering
\includegraphics[width=0.9\linewidth]{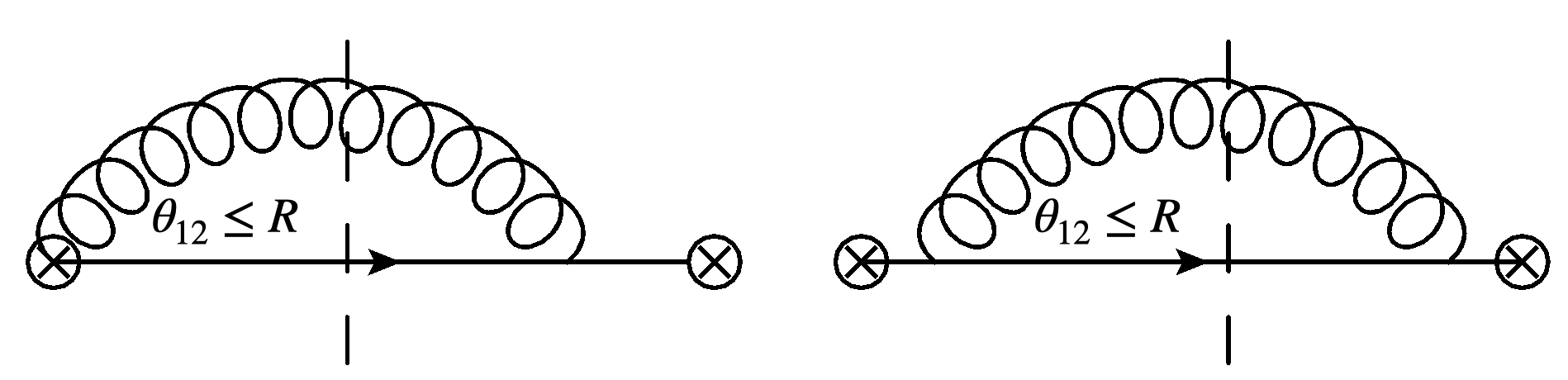}
\caption{Diagrams that contribute to the two sub-jet matching co-efficient.}
\label{fig:C2}
\end{figure}

{\bf Matching for two sub-jets:}
We now derive the explicit leading order matching co-efficient  and color structure for the two sub-jet configuration. 
The general structure for the two sub-jet term takes the form 
\begin{eqnarray}
&&Q_2 = \int d\theta_{12} {\cal C}^{a \bar a kk'i i' }_{q \rightarrow q+g}(z', \omega'_J, \theta_{12}, \mu)  \nonumber \\
&& \text{Tr}\Big[{\bf U}_{lk}(n_1) \boldsymbol{\mathcal{U}}_{ba}(n_2)U_{i j}(\bar n)\rho_M U^{\dagger}_{ji'}(\bar n)\boldsymbol{\mathcal{U}}^{\dagger}_{b \bar a}(n_2){\bf U}^{\dagger}_{k'l}(n_1) \mathcal{M} \Big] \nonumber \\
\end{eqnarray}
where $\mathcal{M}$ is the measurement imposed on the collinear soft radiation. The trace is over the soft(medium) and collinear soft final states. In our case, we are using $kT$ type algorithms which follow sequential recombination of partons and in general can lead to non-trivial clustering effects which complicate the form of factorization. In particular, it was shown that the collinear-soft functions must also know about the energy fraction of the hard partons along with their directions. We are interested in the scenario when the two hard partons are always inside the jet so that the jet axis for anti-kT is $n=(1,0,0,1)$. It was argued in \cite{Becher:2023znt} that for Leading Log(LL) accuracy, which is the limit of strongly ordered cs emissions, the condition for a cs parton to be outside the jet simplifies to the requirement of the cs gluon lying outside a cone of radius R from the jet axis which simplifies the jet algorithm measurement on the collinear soft function. In this paper, we will only work to this order and leave higher order calculations for the future. At this order, the cs functions do not keep track of the energy fractions of the hard collinear quark and gluon. The operator $\boldsymbol{\mathcal{U}} = \mathcal{U}_m\mathcal{U}$ is defined in the adjoint representation since it is sourced by a gluon.  $\theta_{12}$ is the angle between the directions of the two sub-jets which are specified by light-like vectors $n_1$ and $n_2$.  At leading order the Wilson co-efficient takes the form ${\cal C}^{a \bar a kk'i i' }_{q \rightarrow q+g} = t^a_{ik}t^{\bar a}_{k'i'} {\cal C}_{q\rightarrow q+g}$. At this order the diagrams that contribute to ${\cal C}_{1\rightarrow 2}(z', \omega'_J, \theta_{12}, \mu)$ are shown in Fig. \ref{fig:C2}. Note that only diagrams with two partons in the final state contribute to ${\cal C}_{q \rightarrow q+g}$ at leading order so that this Wilson co-efficient starts off at O($\alpha_s$). Here we also require that both these collinear partons should be inside the jet which at this order implies $\theta_{12} \leq R$. 
Since the axis of the jet  points along the $n = (1,0,0,1)$ 
direction, without loss of generality, the two sub-jets can be taken to lie in the $x-z$ plane with $n_1^{\mu}\equiv(1, \sin \theta_1, 0, \cos \theta_1)$ and $n_2^{\mu} \equiv (1, -\sin \theta_2, 0, \cos \theta_2 )$. If we assume that the gluon carries a fraction x of the jet energy, then we also need $(1-x)\sin \theta_1 - x \sin \theta_2 =0$. Working at LL,  where clustering effects can be ignored, $x \rightarrow 0$, so that $\theta_1 \rightarrow 0$ and $\theta_{12} = \theta_1+ \theta_2  \rightarrow \theta_{2}$. This is the soft limit of the gluon where the quark essentially points along the jet axis. 
Computing the diagrams in Fig.\ref{fig:C2} at LL yields 
\begin{eqnarray}
 {\cal C}_{q \rightarrow q+g}(z', \omega'_J,\theta_{12},\mu)   = -\omega'_J\delta(1-z')\frac{\alpha_s}{\pi}\frac{1}{\theta_{12}}\left(l +\frac{3}{2}\right) 
 \label{eq:C2}
\end{eqnarray}
 We see that when integrated over the angle, this result appears has a collinear singularity as $\theta_{12} \rightarrow 0$.

In the limit $\theta_{12}\rightarrow 0$, the two sub-jet configuration reduces to a single sub-jet and hence should ideally be included in $Q_1$. Using the identity $\boldsymbol{\mathcal{U}}^{b a}(n)t^a = {\bf U}^{\dagger}(n)t^b {\bf U}(n)$, in the limit $\theta_{12} \rightarrow 0$ , $n_1 = n_2 =n$ and 
\begin{eqnarray}
&& Q_2 \rightarrow \mathcal{C}(z', \omega'_J, \mu) C_F\text{Tr}\Big[ {\bf U}(n)U(\bar n) \rho_M U^{\dagger}(\bar n){\bf U}^{\dagger}(n) \mathcal{M}\Big] \nonumber
\end{eqnarray}
where $\mathcal{C}(z', \omega'_J, \mu) =\int d\theta_{12} C_{q\rightarrow q+g}(z', \omega'_J, \theta_{12}, \mu)$. To retain only the regime of two resolved subjets, we can therefore subtract this contribution from the two sub-jet operator and add it into the single sub-jet term $Q_1$. We therefore  redefine $Q_2$ as
\begin{eqnarray}
&&Q_2 = \int d\theta_{12} {\cal C}^{a \bar a kk'i i' }_{q \rightarrow q+g}(z', \omega'_J, \theta_{12}, \mu) \nonumber \\
 &&\Bigg\{{\cal S}_2^{ii'kk'a \bar a}(n_1, n_2, \epsilon, \mu) - {\cal S}_2^{ii'kk'a \bar a}(n, n, \epsilon, \mu)\Bigg\}
\end{eqnarray}
This removes the singularity in integral over $\theta_{12}$ as we will see explicitly in the next section. A consequence of this redefinition is that the Wilson co-efficient for the single sub-jet ${\cal C}_{q \rightarrow q}$ which had an IR singularity (Eq.~\ref{eq:CqIR}) gets modified to 
\begin{eqnarray}
 &&{\cal C}_{q \rightarrow q} \rightarrow {\cal C}_{q \rightarrow 1} + C_F {\cal C} \nonumber\\
 &= &  \omega_J'\delta(1-z')\Bigg\{1+\frac{\alpha_s}{2\pi}C_F\Bigg[\frac{l^2}{2}+\frac{3l}{2}+\frac{13}{2}-\frac{3\pi^2}{4}\Bigg]\Bigg\} \nonumber\\
 \label{eq:c1}
 \end{eqnarray}
 which is now IR finite. The result obtained is identical to the matching co-efficient for the quark initiated Fragmenting Jet Function(FJF) in the threshold limit computed in \cite{Dai:2017dpc} although they did not consider multi-sub jet operators. This now yields the Renormalization Group(RG) equations 
 \begin{eqnarray}
\mu \frac{d}{d\mu}\mathcal{C}_{q\to i}(z, \omega_J,\mu) = \int _z^1 \frac{dz'}{z'} \gamma^{i,j}_{\mathcal{C}}\left(\frac{z}{z'},\mu\right)\mathcal{C}_{q\to j}(z',\omega_J,\mu)\nonumber
\end{eqnarray}
\begin{eqnarray}
   \gamma^{q,q}_{\mathcal{C}}(z) &=& \delta(1-z) \frac{\alpha_s C_F}{2\pi}\Big[2 l +3\Big]\nonumber\\
  \gamma^{q+g,q}_{\mathcal{C}}(z) &=& -2\frac{\alpha_s}{\pi}\frac{1}{\theta_{12}}
\end{eqnarray}
We see that  ${\cal C}_{q \rightarrow q+g}$ mixes into ${\cal C}_{q \rightarrow q}$ and the anomalous dimension for ${\cal C}_{q \rightarrow q+g}$ depends on the opening angle between the quark and gluon. 

We see that to ensure the IR finiteness of the two sub-jet Wilson co-efficient, we subtracted the collinear overlap or equivalently the overlap with the single sub-jet configuration. Therefore the redefined two sub-jet function $\tilde {\cal S}_{2}$ has contribution from vacuum and medium induced collinear soft radiation from two separated ($\theta_{12} \sim R$) hard collinear partons. 
The two sub-jet operator( suppressing color indices) can be written as 
\begin{eqnarray}
\hat {\cal S}_{2}(n_1,n_2) = \left(1- \lim_{n_1= n_2}\right){\cal S}_{2}(n_1,n_2)
\end{eqnarray}
Using this as a guide, we can now define the  3 sub-jet operator as
\begin{eqnarray}
&&\hat {\cal S}_{3}(n_1,n_2,n_3)= \nonumber\\
&& \Bigg[1- \sum_{i\neq j \in\{1,2,3\}}\lim_{n_i=n_j} + 2 \lim_{n_1 =n_2= n_3 }\Bigg]{\cal S}_{3}(n_1,n_2,n_3)
\end{eqnarray}
where we have subtracted out the overlap with the single and two sub-jet operators. The sum is over all distinct pairs $i,j$. Note that the third term in the brackets is needed to avoid over subtraction of the $n_1=n_2=n_3$ limit. The higher order sub-jet operators can now be built up in a similar manner. The subtraction of the overlap with lower multiplicity sub-jets ensure the IR finiteness of the Wilson co-efficient for these operators. Computing Wilson co-efficients for these higher multiplicity operators requires a two or higher loop calculation which is beyond the scope of this paper. 
 
\section{sub-jet functions at one loop}
\label{sec:Oneloop}
We now explicitly present the one loop results for the redefined one and two sub-jet collinear soft operators. These functions in the EFT framework describe physics at the scale $Q_{\text{med}} \sim (1-z) \omega_JR$. 
While the results for the single sub-jet were computed within this EFT framework in \cite{Mehtar-Tani:2024smp}, we present them here for completeness before computing the two sub-jet results. 
\subsection{Single sub-jet}

The one loop vacuum evolution term ${\cal S}^{(1)}_{q,0}$ is given in Eq.~\ref{eq:Sq0oneL}. For the higher order terms in Eq.~\ref{eq:SExp}, for $n\geq 1$ we can factorize the medium physics from the jet in rapidity, 
\small
\begin{align}\label{eq:FctAl}
&{\cal S}_{q,n}(\epsilon,\mu)\! =\! (8\pi \alpha_s)^{2n}\Bigg[\prod_{i=1}^{n}\int_{0}^{\infty}\! \!{\rm d}x^-_i\Theta(x^-_i-x_{i+1}^-)\!\int\! \frac{{\rm d}^2\bfk_{i}}{(2\pi)^3}\nonumber \\
&\varphi(\bfk_{i},\mu_\cs,\nu,x^-_i)\Bigg]
{\bf F}_{q,n}(\epsilon; \bfk_{1}, \ldots, \bfk_{n}; x_1^-, \ldots x_n^-;\mu,\nu)\,.
\end{align}
\normalsize
 ${\bfk}_i $ is the transverse momentum exchanged between the jet and the medium in the $i^{th}$ interaction. ${\bf F}_{q,n}$ in the collinear soft jet function with n Glauber gluon exchanges. 
 $\bar x^-$ is the longitudinal location in the medium and dependence of $\varphi$ on $\bar x^-$ is encoded in the density matrix that is assumed to be slowly varying $\bar x^- \gg r^-$. 
where $\varphi(\bfk, \bar x^-)$ is the color density correlator in the medium 
\begin{eqnarray}
\label{eq:Medcorr3}
&&\varphi(\bfk, \bar x^-)= \frac{1}{\bfk^2}\frac{\delta_{ab}}{N_c^2-1}\int \frac{dk^-}{2\pi}\int d^4r e^{i(k^-\frac{r^+}{2}-\bfk \cdot \bfr)}\nonumber\\
&&\text{Tr}\Bigg[\mathcal{O}^{ia}_{s}(r^+, \bfr , r^-+\bar x^-)\rho_M \mathcal{O}^{ib}_{s}(0) \Bigg]
\end{eqnarray}
$\rho_M$ is the medium density matrix and definition of $\mathcal{O}^{ia}_{s}$ is givn in Appendix \ref{app:Ocs}. 
This function obeys the following RG equations at one loop in the two renormalization scales $\mu ,\nu$ corresponding to separation in virtuality and rapidity respectively. 
\begin{align}
&\frac{d  \varphi(\bfk;\nu;\mu)}{d\ln \nu} = \int d^2\bfu\, \mathcal{K}_{\text{BFKL}}(\bfu,\bfk)\,\varphi(\bfk;\nu;\mu),\nonumber\\
&= \frac{\alpha_s N_c}{\pi^2} \int d^2\bfu \Bigg[\frac{\varphi(\bfu;\nu;\mu)}{(\bfu-\bfk)^2}-\frac{\bfk^2\varphi(\bfk;\nu;\mu) }{2\bfu^2(\bfu-\bfk)^2}\Bigg],\nonumber\\
&\frac{d\varphi(\bfk;\nu;\mu)}{d\ln \mu} =  -\frac{\alpha_s \beta_0}{\pi} \varphi(\bfk;\nu;\mu). 
\label{eq:BRG}
\end{align}
The one loop result for the single interaction collinear soft function was computed in \cite{Mehtar-Tani:2024smp,Mehtar-Tani:2025xxd} and reads
\begin{align}\label{eq:F1q}
&{\bf F}_{q,1}^{(1)}(\epsilon,{\bfk},x^-,\mu)=\frac{\alpha_s(N_c^2-1)}{4\pi^2}\mu^{2\epsilon} \int \frac{{\rm d}^{2-2\epsilon}\bfq}{\bfk^2 \bfq^2} \frac{2 {\bfk} \cdot {\bfq}}{({\bfq}+{\bfk})^2}  \nonumber \\
& \int\frac{{\rm d}q^-}{q^-}\Theta\left(|\bfq|- \frac{q^-R}{2}\right)\Big[\delta(\epsilon)-\delta(q^--\epsilon)\Big] \nonumber \\
&\times\left\{1- \cos \left[\frac{({\bfq}+{\bfk})^2}{2q^-}x^-\right]\right\}\,,
\end{align}
which agrees with GLV results in Refs.~\cite{Gyulassy:2000fs,Gyulassy:2000er,Wiedemann:2000za}.
In Section \ref{sec:cfact}, we will also derive its RG equation through consistency of factorization.

\subsection{Two sub-jets}
\begin{figure}
\centering
\includegraphics[width=0.5\linewidth]{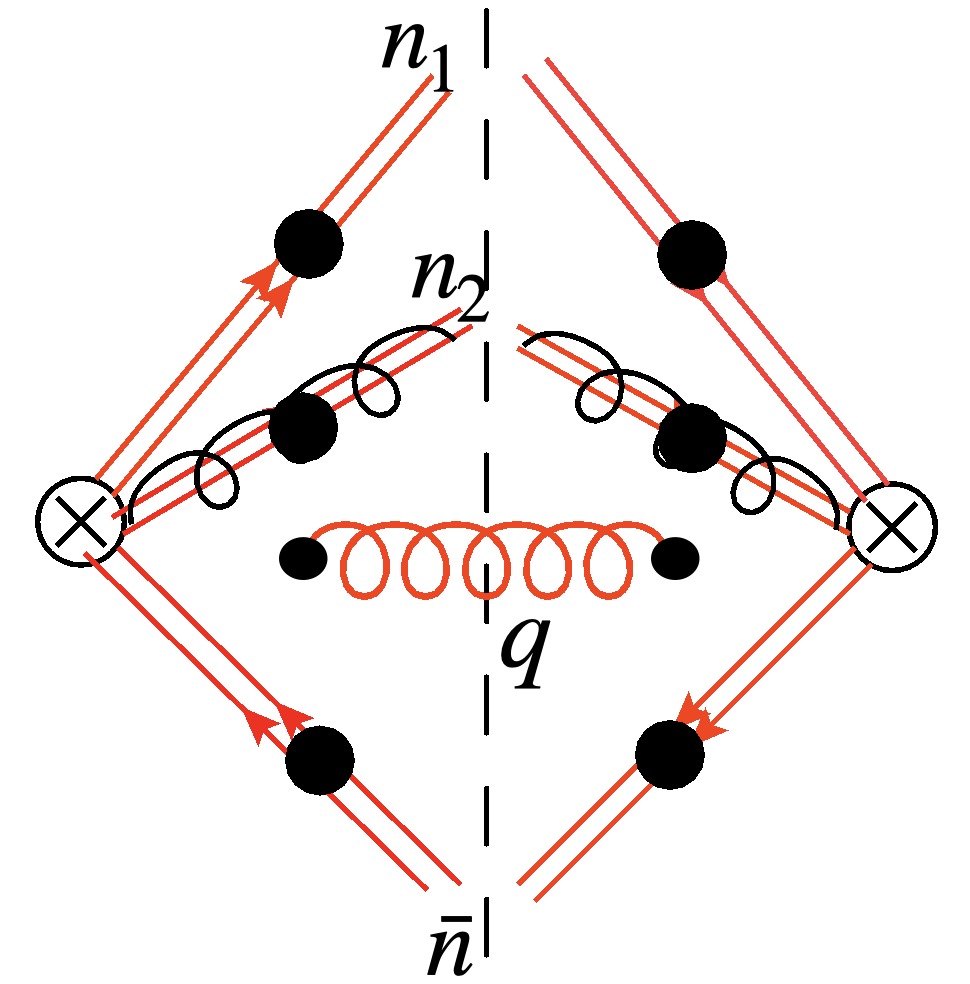}
\caption{Real gluon emission diagram for vacuum evolution of two sub-jets. Lines along $n_1$ and $n_2$ are denote the cs Wilson lines in the fundamental and adjoint representation respectively. The black vertices denote the possible attachments of the Feynman diagrams.}
\label{fig:s2vac}
\end{figure}

As with the single sub-jet, the two sub-jet operator can be expanded order by order in the number of interactions with the medium. 
\begin{eqnarray}
    &&Q_2 =  \int d\theta_{12}C^{a \bar a ii'kk'}_{q\rightarrow q+g}(z', \theta_{12},\mu)\nonumber \\ 
    & &\sum_{k=0}^{\infty}\Bigg\{{\cal S}_{2,k}^{ii'kk'a \bar a}(n_1, n_2, \epsilon, \mu) - {\cal S}_{2,k}^{ii'kk'a \bar a}(n, n, \epsilon, \mu)\Bigg\} \nonumber \\
    &\equiv & \int d\theta_{12}{\cal C}^{a \bar a ii'kk'}_{q\rightarrow q+g}(z', \theta_{12},\mu)\sum_{k=0}^{\infty} \tilde {\cal S}_{2,k}^{ii'kk'a \bar a}(n_1,n_2,\epsilon,\mu) \nonumber \\
    \label{eq:2sbjexp}
\end{eqnarray}
{\bf Vacuum evolution :}
To begin with, consider the the first term in this series with no medium interactions, namely vacuum evolution. This corresponds to ignoring the $H_{cs-s}$ interaction term in the Hamiltonian, Eq.~\ref{eq:H2} and setting the $U_{{\bf M}}$ operator to identity. At LL, which is the limit of strongly ordered emissions, ${\cal C}^{a \bar a ii'kk'}_{q\rightarrow q+g} \rightarrow t^a_{ik}t^{\bar a}_{k'i'}{\cal C}^{\text{LL}}_{q \rightarrow q+g}$ and  $n_1 \rightarrow n$ coincides with the jet axis so that \small ${\cal S}_{2,0}^{ii'kk'a \bar a} $ is now reduces to $ \text{Tr}\Big[ U_{lk}(n) \mathcal{U}_{ba}(n_2)U_{i j}(\bar n)\rho_M U^{\dagger}_{ji'}(\bar n)\mathcal{U}^{\dagger}_{b \bar a}(n_2) U^{\dagger}_{k'l}(n) \mathcal{M} \Big] $. \normalsize
At one loop, the virtual diagrams vanish in dimensional regularization. All the the real diagrams are shown in Fig. \ref{fig:s2vac}. There are three non-zero contributions corresponding to the cs gluon sourced by  $n- \bar n,$ $n_2-\bar n$ and $n_2- \bar n$. The contribution when the cs gluon is inside the jet is also scaleless and hence goes to zero. So we only consider the case when the cs gluon is emitted outside the jet.
This give us 
\begin{eqnarray}
&&t^a_{ik}t^{\bar a}_{k'i'}\tilde {\cal S}^{(1)ii'kk'a \bar a}_{2,0}= g^2C_F\left(\frac{\mu^2 e^{\gamma_E}}{4\pi}\right)^{\epsilon} \frac{C_A}{2}\frac{1}{\epsilon_L}\int \frac{d^{2-2\epsilon}\bfq}{(2\pi)^{d-1}}\nonumber \\
&&\Theta\left( \frac{|\bfq|}{\epsilon_L}-\frac{R}{2}\right)\left(-\frac{n \cdot \bar n}{n \cdot q \bar n \cdot q}+ \frac{n_2 \cdot \bar n}{n_2 \cdot q \bar n \cdot q} + \frac{n_2 \cdot n}{n_2 \cdot q  n \cdot q} \right)\nonumber
\end{eqnarray}
where $q^- = \bar n \cdot q =\epsilon_L$, the contribution to the energy loss. We can also write $n_2 \cdot \bar n = 1+ \cos \theta_{12} \approx 2 -\theta_{12}^2/2 \approx 2$, since $\theta_{12}$ is bounded by $R \ll 1$. Similarly $n_2 \cdot q \approx  |\bfq|^2/\epsilon_L+ \theta_{12}^2/(4\epsilon_L)+|\bfq|\theta_{12} \cos \phi_q$. We can further rescale $\bfq $ by $\epsilon_L R/2$ to write the result in terms of $\hat \theta_{12} = 2\theta_{12}/R$
\begin{eqnarray}
&&t^a_{ik}t^{\bar a}_{k'i'}\tilde {\cal S}^{(1)ii'kk'a \bar a}_{2,0}= 2\frac{\alpha_s C_F}{2\pi^2} \frac{N_c}{2}\left(\frac{4\mu^2e^{\gamma_E}}{4\pi R^2 \epsilon^2_L}\right)^{\epsilon}\frac{1}{\epsilon_L}\nonumber \\
&&\int \frac{d^{2-2\epsilon}\bfq}{\bfq^2} \Theta\left( |\bfq|-1\right)\left(\frac{2|\bfq|\hat \theta_{12}\cos \phi_q}{\bfq^2+ \hat \theta_{12}^2/4+ |\bfq|\hat \theta_{12}\cos \phi_q}\right)\nonumber
\end{eqnarray}
We note that the integral over $\bfq$ is both UV and IR finite. Writing $\epsilon_L= (1- \tilde z) \omega_J'$, We can explicitly evaluate this to give 
\small
\begin{eqnarray}
&&t^a_{ik}t^{\bar a}_{k'i'}\tilde {\cal S}^{(1)ii'kk'a \bar a}_{2,0}= -2\frac{\alpha_s C_F}{\pi \omega_J'} \frac{N_c}{2}\frac{\left(\frac{4\mu^2e^{\gamma_E}}{4\pi R^2 }\right)^{\epsilon} }{(1-\tilde z)^{1+2\epsilon}}\ln \left(1-\frac{\theta_{12}^2}{R^2}\right)\nonumber\\
&=& \frac{\alpha_s C_F}{\pi \omega_J'} \frac{N_c}{2}\left(-\frac{2}{[1-\tilde z]_+}+ \delta(1-\tilde z)\left(\frac{1}{\epsilon}+ l\right)\right) \ln \left(1-\frac{\theta_{12}^2}{R^2}\right) \nonumber
\end{eqnarray}
\normalsize
For small opening angles $\theta_{12} \ll R$, this function scales as $\theta_{12}^2$, thus when convolved with the Wilson co-efficient $C_{1\rightarrow 2}$, it regulates the collinear singularity. This leads us to the RG evolution equation for $S_2$ at leading log
\begin{eqnarray}
&&\mu \frac{d}{d\mu}t^a_{ik}t^{\bar a}_{k'i'}\tilde {\cal S}^{ii'kk'a \bar a}_{2,0}(z, \theta_{12}, \mu) \nonumber\\
&= & \int _z^1 \frac{dz'}{z'} \gamma^{2,1}_{{\cal S}}\left(\frac{z}{z'},\mu, \theta_{12} \right){\cal S}_{q,0}(z',\omega_J,\mu,\theta_{12})\nonumber
\end{eqnarray}
\begin{eqnarray}
 \gamma^{2,1}_{{\cal S}}(z)= \frac{\alpha_sC_F}{\pi}N_c\ln 
\left(1- \frac{\theta_{12}^2}{R}\right)\delta(1-z)
\end{eqnarray}
Once again we see that the two sub-jet collinear soft function mixes into the one sub-jet operator. While we have RG scale independence to $O(\alpha_s)$, to do the same at O($\alpha_s^2$),  we also need the two loop results for $C_{q \rightarrow q}$ and ${\cal S}_{q,0}$ which is beyond the scope of this paper. 

{\bf Medium evolution:}
We now consider the two sub-jet function with a single interaction with the medium. This is the second term in the series in Eq.~\ref{eq:2sbjexp}. 
As with the single sub-jet scenario, the function $\tilde S_{2,1}$ can be factorized as
\small
\begin{eqnarray}
&& \tilde {\cal S}^{a\bar a ii'kk'}_{2,1}(n_1,n_2,\epsilon_L)=|C_G|^2\, \int d \bar x^-\int \frac{d^2\bfk}{(2\pi)^3 } \nonumber \\
&&\textbf{F}_{2,1}^{a\bar a ii'kk'}(n_1,n_2,\epsilon_L;\bfk, \bar x^-)\varphi(\bfk, \bar x^-).
\end{eqnarray}
\normalsize
where $\textbf{F}_{2,1}^{a\bar a ii'kk'} = \textbf{F}_{2,1}^{a\bar a ii'kk',r}-\textbf{F}_{2,1}^{a\bar a ii'kk',v}$ and $\textbf{F}_{2,1}^{r}$($\textbf{F}_{2,1}^{v}$) is the collinear soft function with Glauber exchange on opposite(same) side of the cut. $\bfk $ is the transverse momentum exchanged between the jet and the medium. The function $\varphi(\bfk,  \bar x^-)$ is identical to that which appears in the single sub-jet and was defined in Eq.~\ref{eq:Medcorr3}. 
The function $\textbf{F}_{2,1}^{r}$ is defined as
\begin{widetext}
    \begin{align}
&\textbf{F}^{a\bar a ii'kk',r}_{2,1}(\underline{n},\epsilon_L, \bfk ,\bar x^-) = \frac{1}{2N_c}\frac{1}{\bfk^2}\nonumber\\ 
& \sum_{X}\tr\Big[\langle 0|\sum_{r=1,2}e^{-i\frac{\bar x^-}{2}\left(\mathcal{P}_{r,+}-\bfk \cdot n_r\right)}\Big[\delta(\mathcal{P}^-)\delta^2(\mathcal{P}_{r\perp}-\bfk)\mathcal{O}_{n_r}^{q,e}(0)+ \frac{1}{({\mathcal{P}}_{r\perp}-\bfk)^2}\mathcal{O}_{n_r,\cs}^{ce}(0)t^c\Big]U^{\dagger}_{ji'}(\bar n)\mathcal{U}^{\dagger}_{b \bar a}(n_2) U^{\dagger}_{k'l}(n) \mathcal{M}'|X\rangle \nonumber\\
&\langle X|\sum_{s=1,2}e^{i\left(\frac{\bar x^-}{2}\mathcal{P}_{s,+}-\bfk \cdot n_s\right)}\Big[\delta(\mathcal{P}^-)\delta^2(\mathcal{P}_{s\perp}+\bfk)\mathcal{O}_{n_s}^{q,f}(0)+ \frac{1}{({\mathcal{P}}_{s\perp}-\bfk)^2}\mathcal{O}_{n_s,\cs}^{ce}(0)t^c\Big]U_{lk}(n) \mathcal{U}_{ba}(n_2)U_{i j}(\bar n)|0\rangle \Big] - \Big\{ n_1,n_2 \rightarrow n\Big\}\,.
\end{align}
\end{widetext}
 The $\mathcal{P}_{r\perp}$ extracts the transverse momentum with respect to the axis $n_r$ of the operator that it acts on. The operators $\mathcal{O}_{n_s,\cs}^{ce}, \mathcal{O}_{n_s}^{q,f}$ which correspond to medium induced radiation and gluon rescattering respectively, are derived from the Glauber SCET framework \cite{Rothstein:2016bsq} and defined in Appendix \ref{app:Ocs}. The operator with Glauber operator insertions on the same side of the cut $\textbf{F}_{2,1}^{v}$ can be defined in the same way. In this paper, we compute the collinear soft function $\left(\textbf{F}_{2,1}^{r}-\textbf{F}_{2,1}^{v}\right) $ at Leading Log which corresponds to the limit $n_1=n$. The tree level result vanishes while the one loop result has three types of contributions: 1. Broadening of vacuum radiation 2. Medium induced radiation 3. Interference between vacuum and medium induced radiation. We now consider each contribution in turn, working in the Feynman gauge. 

\begin{figure}
\centering
\includegraphics[width=0.9\linewidth]{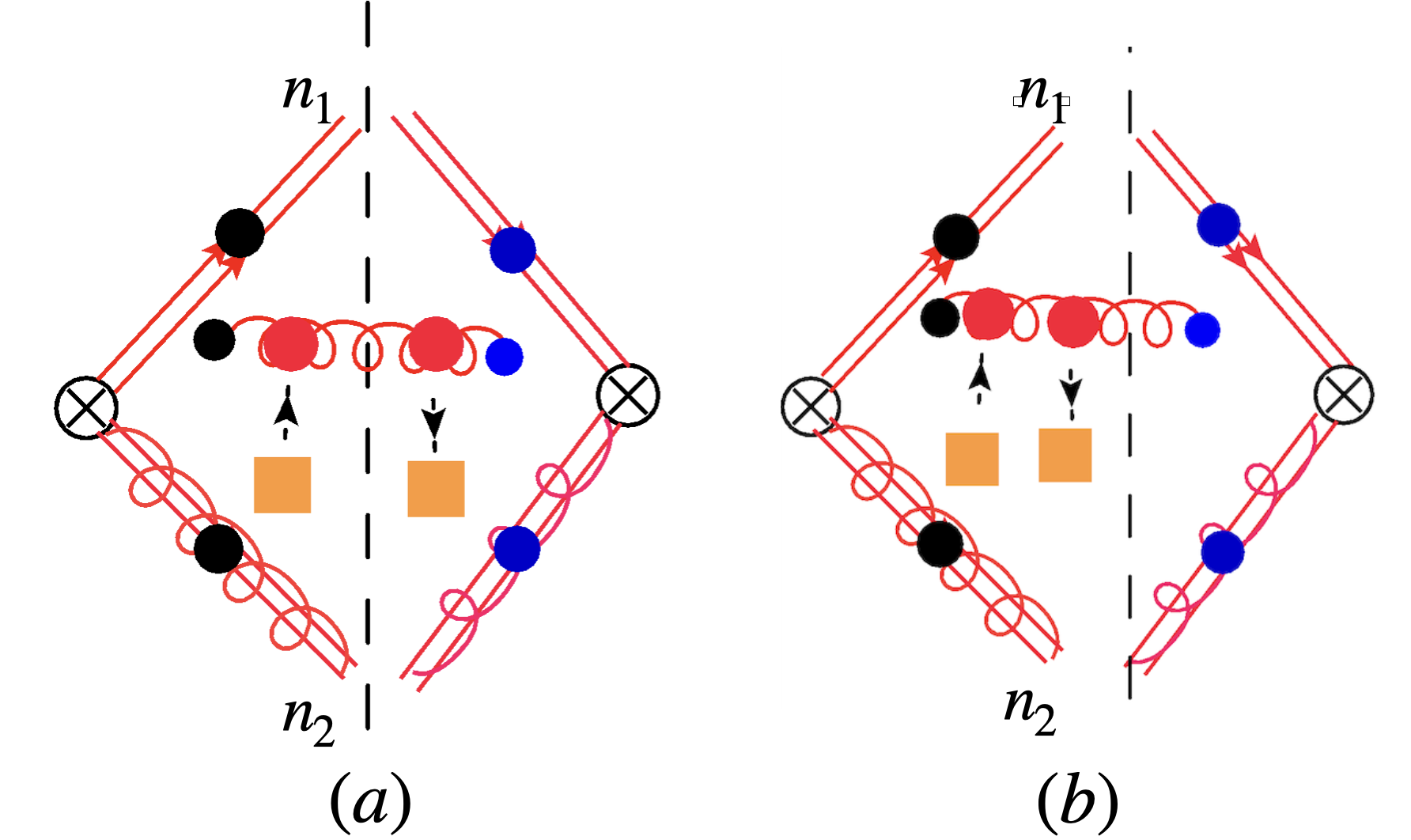}
\caption{The lines along $n_1$($n_2$) denote cs Wilson lines in the fundamental(adjoint) representation. The black and blue vertices denote the possible attachments for contributing Feynman diagrams.}
\label{fig:vbv}
\end{figure}

{\bf Vacuum radiation broadening:}
We first consider the contribution from broadening of vacuum radiation sourced by the two collinear-soft Wilson lines. At LL, the fundamental Wilson line $U(n_1)$ is along the jet axis i.e. $n_1 =n$. The SCET Feynman rule for the n collinear-soft gluon  scattering off the medium is shown in Eq.\ref{eq:gsc}.
\begin{align}
\begin{minipage}{3cm}\includegraphics[width=\textwidth]{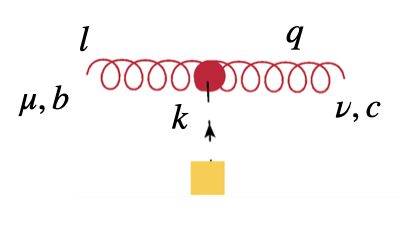} \end{minipage}
=&-8\pi \alpha_s if^{abc}\Big[\bar{n}\cdot l g_{\perp}^{\mu \nu}-\bar{n}^{\mu} {\bf l}^{\nu} \nonumber\\
&-\bar{n}^{\nu}\bfq^{\mu}+\frac{\bfq\cdot {\bf l} \bar{n}^{\mu}\bar{n}^{\nu}}{\bar{n}\cdot q} \Big]. 
\label{eq:gsc}
\end{align}
The diagrams corresponding to real gluon emission are shown in Fig.~\ref{fig:vbv}. As explained in \cite{Rothstein:2016bsq}, the Glauber limit of the cs gluon before scattering off the medium is already included in the operator $\mathcal{O}_{cs}$ for medium induced radiation. Hence to prevent double counting, we need to implement a Glauber zero bin subtraction. In this case, considering the real and virtual diagrams, we get the result  
\begin{eqnarray}
 && 5(a)+5(b)+ \text{virtual}= \frac{\alpha_s}{2\pi^2}\int \frac{dq^-}{q^-}\int d^2\bfq \mathcal{M}^{(1)}(q) \nonumber\\
&&\Bigg[\frac{N_c^2\delta^{a\bar a}\delta_{kk'}\delta_{ii'}}{(\bfq_2-\bfk)^2} + 2iN_cf^{a\bar a c}t^c_{kk'}\delta_{ii'}\frac{(\bfq-\bfk)\cdot (\bfq_2-\bfk)}{(\bfq-\bfk)^2(\bfq_2-\bfk)^2}\nonumber\\
& & -\bfk \rightarrow 0  \Bigg] - \Big\{n_2 \rightarrow n \Big\}
\end{eqnarray}
where $\mathcal{M}^{(1)}(q)=\Theta \left(\frac{|\bfq|}{q^-}-\frac{R}{2}\right)\Big[\delta(\epsilon_L)-\delta(\epsilon_L-q^-)\Big]$ and 
${\bfq}$(${\bfq}_2$) is the transverse momentum of the emitted gluon with respect to the axis n($n_2$). 

\begin{figure}
\centering
\includegraphics[width=0.9\linewidth]{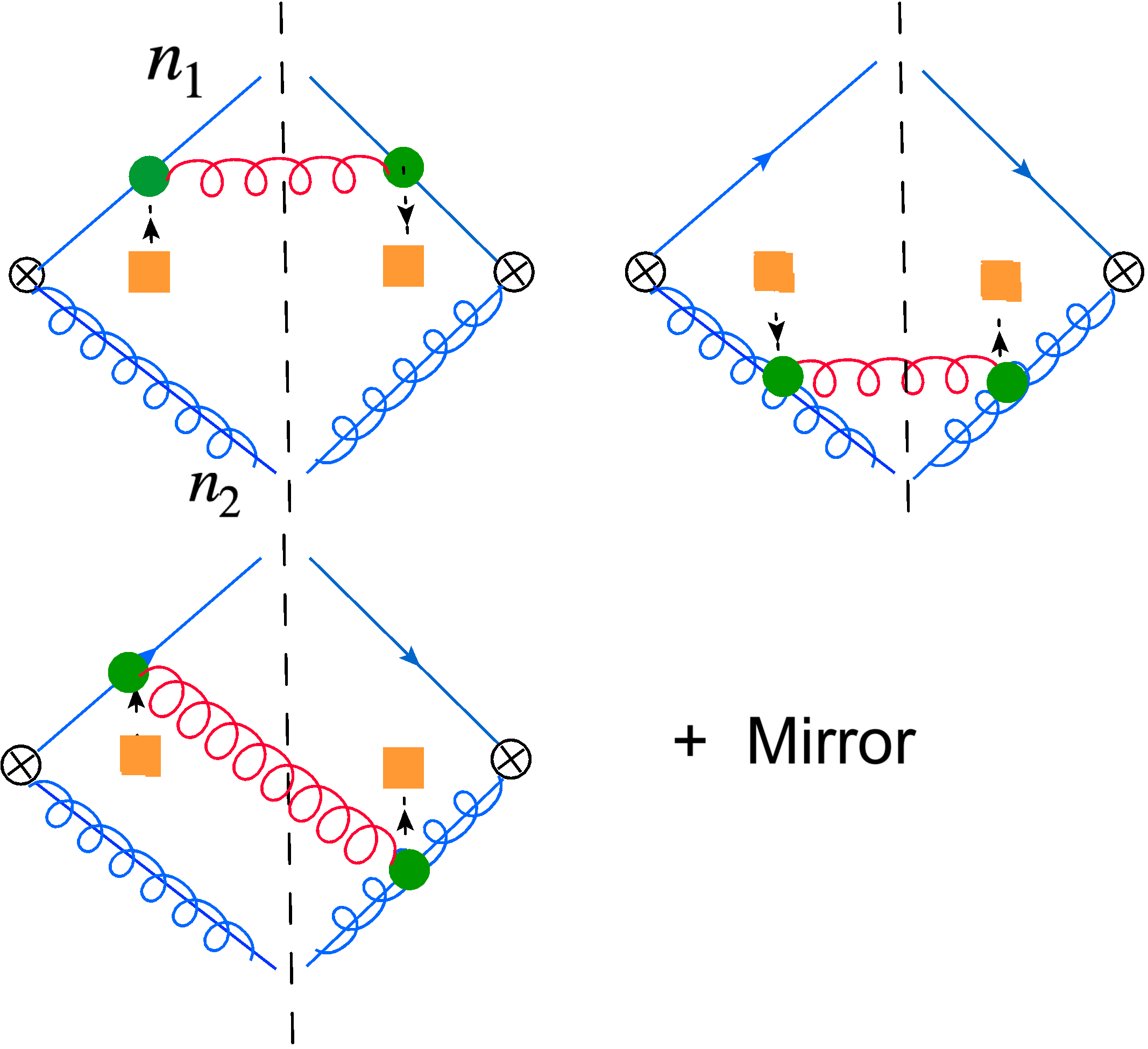}
\caption{Blue lines along $n_1$($n_2$) denote the medium induced radiation operator ${\bf U}_M$ (Eq.~\ref{eq:CSMed}) in the fundamental(adjoint representation)}
\label{fig:thetac}
\end{figure}

{\bf Medium Induced radiation:} Next we consider only the contribution from radiation induced through interaction with the medium.
 The Feynman rule that we need is shown in Eq. \ref{eq:mir} is derived from the operator $\mathcal{O}_{cs}$. The corresponding diagrams Feynman diagrams are shown in Fig.\ref{fig:thetac}.
\begin{align}
\begin{minipage}{3cm}\includegraphics[width=\textwidth]{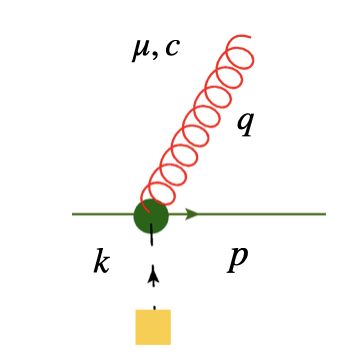} \end{minipage}=&8\pi \alpha_s igf^{abc}\Bigg[\!\!2\bfk^{\mu}-\bfq^{\mu}-\frac{\bar{n}^{\mu}}{2}n\cdot q \nonumber\\
&+\bar{n}\cdot q \frac{n^{\mu}}{2}-n^{\mu}\frac{(\bfk-\bfq)^2}{n\cdot q}+\bar{n}^{\mu}\frac{\bfk^2}{\bar{n}\cdot q}\!\Bigg] \nonumber 
\label{eq:mir}
\end{align}

\begin{figure}
\centering
\includegraphics[width=\linewidth]{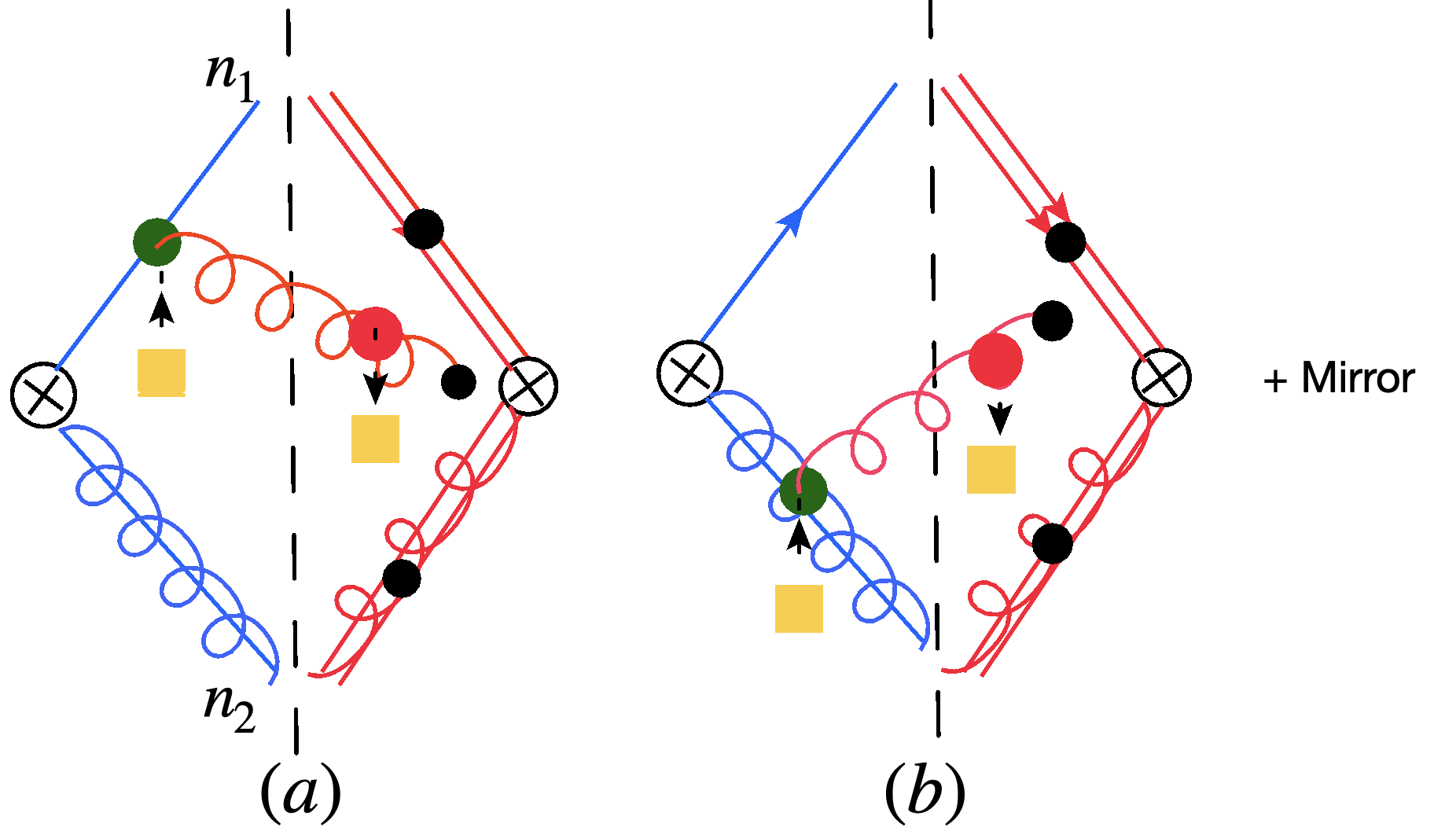}
\caption{Real emission gluon diagrams with interference between medium induced radiation(green vertices) and vacuum radiation(black vertices). }
\label{fig:intmb}
\end{figure}

Combining all the terms including the virtual diagrams, the result can again be decomposed into the same color structures as  
\begin{eqnarray}
&&\text{Fig}.6+ \text{virtual} =  \frac{\alpha_s}{2\pi^2}\int \frac{dq^-}{q^-}\int d^2\bfq \mathcal{M}^{(1)}(q) \nonumber\\
&& \Bigg[\frac{N_c\delta^{a\bar a}\delta_{kk'}\delta_{ii'} \bfk^2}{\bfq_2^2(\bfq_2-\bfk)^2}+  2if^{a\bar a c}t^c_{kk'}\delta_{ii'}\left(\frac{\bfq -\bfk}{(\bfk-\bfq)^2}-\frac{\bfq}{\bfq^2}\right) \nonumber\\
&&\cdot\left(\frac{\bfq_2 -\bfk}{(\bfk-\bfq_2)^2}-\frac{\bfq_2}{\bfq_2^2}\right) \times \cos\left( \frac{(\bfq-\bfk)^2}{2q^-}x^--\frac{(\bfq_2-\bfk)^2}{2q^-}x^-\right) \nonumber \\
& & -\bfk \rightarrow 0  \Bigg] - \Big\{n_2 \rightarrow n \Big\}
\end{eqnarray}
{\bf Interference between vacuum broadening and Medium induced radiation}
The final set of diagrams corresponding to the interference terms are as shown in Fig.\ref{fig:intmb} and yield the result
\small
\begin{eqnarray}
&&\text{Fig}.7 +\text{virtual} =  \frac{\alpha_s}{2\pi^2}\int \frac{dq^-}{q^-}\int d^2\bfq \mathcal{M}^{(1)}(q) \nonumber\\
& & \Bigg[\frac{N_c\delta^{a\bar a}\delta_{kk'}\delta_{ii'} \bfk \cdot \bfq_2}{\bfq_2^2(\bfq_2-\bfk)^2}\cos \left(\frac{(\bfq_2-\bfk)^2}{2q^-}x^-\right) \nonumber \\
&+& 2if^{a\bar a c}t^c_{kk'}\delta_{ii'}\Bigg\{\Bigg[\frac{(\bfq -\bfk)\cdot(\bfq_2-\bfk)}{(\bfk-\bfq)^2(\bfk-\bfq_2)^2}-\frac{\bfq \cdot( \bfq_2-\bfk)}{\bfq^2(\bfk-\bfq_2)^2}\Bigg] \nonumber\\
&& \cos\left( \frac{(\bfq-\bfk)^2}{2q^-}x^-\right)+\Bigg[\frac{(\bfq_2 -\bfk)\cdot(\bfq-\bfk)}{(\bfk-\bfq_2)^2(\bfk-\bfq)^2} -\frac{\bfq_2 \cdot( \bfq-\bfk)}{\bfq_2^2(\bfk-\bfq)^2}\Bigg]\nonumber\\
 &&\cos\left( \frac{(\bfq_2-\bfk)^2}{2q^-}x^-\right)-\bfk \rightarrow 0  \Bigg\}\Bigg] - \Big\{n_2 \rightarrow n \Big\}
\end{eqnarray}
\normalsize
We can now combine all the contributions to write our final result for the two sub-jet single medium interaction collinear soft function at one loop.
\small
\begin{eqnarray}
&& \textbf{F}_{2,1}^{a \bar a ii'kk'(1)}=  \frac{\alpha_s}{2\pi^2 \bfk^2}\int \frac{dq^-}{q^-}\int d^2\bfq \mathcal{M}^{(1)}(q) \nonumber\\
&&\Bigg[\frac{N_c\delta^{a\bar a}\delta_{kk'}\delta_{ii'} \bfk \cdot \bfq_2}{\bfq_2^2(\bfq_2-\bfk)^2}\Bigg[1- \cos \left(\frac{(\bfq_2-\bfk)^2}{2q^-}x^-\right)\Bigg] \nonumber\\
&+& 2if^{a\bar a c}t^c_{kk'}\delta_{ii'}\Bigg\{ {\bf L}\cdot\frac{(\bfq_2-\bfk)}{(\bfk-\bfq_2)^2}\Bigg [1- \cos\left( \frac{(\bfq-\bfk)^2}{2q^-}x^-\right)\Bigg]
\nonumber\\
&+&\bar {\bf L}\cdot\frac{(\bfq-\bfk)}{(\bfk-\bfq)^2} \Bigg[1-\cos\left( \frac{(\bfq_2-\bfk)^2}{2q^-}x^-\right)\Bigg]\nonumber\\
&+&{\bf L} \cdot \bar{\bf L} \Bigg[1-\cos\left( \frac{(\bfq-\bfk)^2}{2q^-}x^--\frac{(\bfq_2-\bfk)^2}{2q^-}x^-\right)\Bigg] \Bigg\}\Bigg]\nonumber\\
&- &\Big\{n_2 \rightarrow n \Big\}
\end{eqnarray}
\normalsize
where we have defined  
\begin{eqnarray}
{\bf L} &=& \frac{\bfq -\bfk}{(\bfk-\bfq)^2}-\frac{\bfq}{\bfq^2}, \ \ \ {\bf \bar L} =\frac{\bfq_2 -\bfk}{(\bfk-\bfq_2)^2}-\frac{\bfq_2}{\bfq_2^2} 
\end{eqnarray}
In order to compare with earlier results in literature we compute the leading fixed order result $ t^a_{ik}t^{\bar a}_{k'i'}{\bf F}_{1,2}^{a\bar a ii'kk'(1)}$ which leads to 
\small
\begin{eqnarray}
 && t^a_{ik}t^{\bar a}_{k'i'}{\bf F}_{1,2}^{a\bar a ii'kk'(1)} = N_cC_F\frac{\alpha_s}{2\pi^2}\int \frac{dq^-}{q^-}\int d^2\bfq \mathcal{M}^{(1)}(q)\nonumber \\
 &\times& \Bigg\{-\Bigg(1- \cos\left( \frac{(\bfq_2-\bfk)^2}{2q^-}x^-\right){\bf K} \cdot {\bf \bar L}\nonumber\\
 &+&\Bigg(1- \cos\left( \frac{(\bfq-\bfk)^2}{2q^-}x^-\right){\bf K} \cdot {\bf L} \nonumber\\
 &+&  \Bigg[1-\cos\left( \frac{(\bfq-\bfk)^2}{2q^-}x^--\frac{(\bfq_2-\bfk)^2}{2q^-}x^-\right)\Bigg] {\bf L} \cdot {\bf \bar L}\Bigg\}
 \label{eq:Dec}
\end{eqnarray}
\normalsize
where we have defined  
\begin{eqnarray}
 {\bf K} =  \frac{\bfq -\bfk}{(\bfk-\bfq)^2}-\frac{\bfq_2 -\bf k}{(\bfk-\bfq_2)^2}
\end{eqnarray}
This result  derived  using the Feynman rules of the Effective SCET Lagrangian agrees with the perturbative result obtained in the Light Cone gauge \cite{Mehtar-Tani:2011lic} using full QCD. This is a powerful cross check of our EFT framework.
The first two terms combined $\rightarrow 0$ as $n_2 \rightarrow n \equiv \theta_{12} \rightarrow 0$. Similarly the third term also goes to 0 as $n_2 \rightarrow n$.  Therefore this result regulates the divergence as $\theta_{12} \rightarrow 0$ in the the Wilson coefficient ${\cal C}_{q\rightarrow q+g}$ in Eq.~\ref{eq:C2}. 

The phase factor corresponding to the first two terms of this result proportional to ${\bf K} \cdot {\bf \bar L}$ and ${\bf K} \cdot {\bf L}$ captures the LPM effect, the interference between vacuum and medium induced radiation. This leads to a suppression if the lifetime of the the emitted gluon $2q^-/(\bfq_i -\bfk)^2$ is much smaller than the medium size $x^-$. For collinear soft radiation, in a dilute medium,  $\bfq, \bfq_2 \sim m_D$ while $q^- \sim m_D/R$ so that the LPM effect is controlled by the emergent power counting parameter $\lambda_m = m_D RL= \sqrt{\hat q L}R L $, where we have introduced the jet quenching parameter $\hat q \sim m_D^2/L$ for a dilute medium. 

We can simplify the phase factor of the third term proportional to $ {\bf L} \cdot {\bf \bar L}$ as 
\begin{eqnarray}
   \phi & =&\frac{(\bfq-\bfk)^2}{2q^-}x^--\frac{(\bfq_2-\bfk)^2}{2q^-}x^- = x^-\frac{\bfq+ \bfq_2 -2\bfk}{2} \cdot \delta {\bf n}, \nonumber \\
    &&\text{where} \ \ \delta{\bf n}= \frac{\bfq-\bfq_2}{q^-}
\end{eqnarray}
and for small angles $|\delta {\bf n}| = \theta_{12}$. Given the scaling $\bfq \sim \bfq_2\sim \bfk \sim m_D$,  $|\phi| \sim \theta_{12}x^- m_D \equiv r_{\perp} m_D$ where $r_{\perp}$ is the transverse size of the q-g dipole. Once we integrate over $x^-$, $r_{\perp} \sim \theta_{12} L$ .Similarly integrating over $\theta_{12}$ after convolution with the Wilson co-efficient ${\cal C}_2$ implies that $\theta_{12} \sim R$. Therefore, the phase factor $\phi$ will scale as $R /\theta_c \sim Rm_D L  \sim \sqrt{\hat q L}R L $. Thus, once again, we have the same emergent power counting parameter $\lambda_m = \sqrt{\hat q L}R L $ that controls color decoherence. 

We can therefore consider three regimes controlled by $\lambda_m$.  
For $\lambda_m \ll 1$, there is a strong suppression of radiation both due to the LPM effect and inability of the medium to resolve multiple sub-jets. In this case, medium effects are irrelevant and the jet essentially evolves in vacuum. Considering a typical value of $R \sim 0.3$ for our inclusive observable, this would be relevant for small $L \sim 1$ fm and dilute ($\hat q \leq 0.1 \text{Gev}^2/$ fm) systems. For $\lambda_m \sim 1$, interference effects are important but there is still $O(1)$ medium contribution compared to the vacuum. This is relevant for intermediate systems $L \sim 3$ fm , $\hat q \sim 0.5\text{GeV}^2/$fm. Finally for $\lambda_m \gg 1$, we can ignore all the $\cos$ terms  in Eq.~\ref{eq:Dec} since they vanish after integration. This corresponds to the regime where every hard collinear parton created during vacuum evolution inside the jet acts as an independent source of medium induced cs radiation. Likewise, all cs radiation created through vacuum evolution goes on-shell before rescattering off the medium. This would be true for large $L \geq 5 $ fm and dense $\hat q \geq 1 \text{GeV}^2/$fm. For more exclusive jet substructure observables, such as the two point energy correlator, as we sweep across the angular separation $\chi$ inside the jet, we will sequentially pass through all three regimes. So the relevance of each regime also depends on the angular scale probed by the observable, which is seen in our case through the dependence on R.

 We see that the lowest order non-zero result for function $\textbf{F}_{r,2}^{a\bar a ii'kk'(1)}-\textbf{F}_{v,2}^{a\bar a ii'kk'(1)}$ at O$(\alpha_s)$ Eq.~ref{} is UV finite. Hence to see the RG equation for this function requires a two loop calculation. In the next section, we will comment on how the RG evolution of atleast some of the  functions considered in this paper can be obtained through consistency of factorization. 

 \section{Consistency of factorization}
 \label{sec:cfact}

 Working order by order in $\alpha_s$, we demand that the cross section should be independent of the factorization scales $\mu,  \nu$, which allows us to infer the RG equations for the functions considered in this paper. We first consider consistency of factorization in the rapidity renormalization scale $\nu$. Given the RG evolution for the $\varphi$ function in Eq.~\ref{eq:BRG}, we infer that the one and two sub-jet single medium interaction collinear soft functions ${\bf F}^{r} \in \{ {\bf F}_{q,1},{\bf F}^{a \bar aii'kk'}_{2,1}\}$ must obey the same equation in rapidity
 \begin{eqnarray}
&\frac{d \textbf{F}^{r}(\bfk;\nu;\mu)}{d\ln \nu} = -\int d^2\bfu\, \mathcal{K}_{\text{BFKL}}(\bfu,\bfk)\,\textbf{F}^{r}(\bfk;\nu;\mu)
\label{eq:FRG}
\end{eqnarray}
 indicating that there is no mixing between operators in the rapidity sector. To see this explicitly requires a 2 loop calculation, but the EFT allows to access logarithmic higher order corrections through consistency alone.
 Similarly, imposing consistency of factorization upto $O(\alpha_s)$ anomalous dimensions in the scale $\mu$ allows us to infer
 \begin{eqnarray}
  &&\frac{d}{d \ln \mu} {\bf F}_{q,1}(\bfk;\mu;\nu)= \int_{z}^1 \frac{dz'}{z'}\Bigg[\gamma_{{\bf F}}^{q\rightarrow q}\left(\frac{z}{z'},\mu\right) {\bf F}_{q,1}(z', \mu) \nonumber\\
  &+&\int d\theta_{12}\gamma_{\bf F}^{q \rightarrow q+g}\left(\frac{z}{z'},\mu,\theta_{12}\right) t^{a}_{ik}t^{\bar a}_{k'i'}{\bf F}^{a\bar a ii'kk'}_{2,1}(z', \mu,\theta_{12})\Bigg]\nonumber\\
 \end{eqnarray}
 where 
 \begin{eqnarray}
     \gamma_{{\bf F}}^{q\rightarrow q}\left(z,\mu\right) &=& -\delta(1-z)\frac{2\alpha_sC_F}{2\pi}l +\frac{\alpha_sC_F}{2\pi}\frac{4}{(1-z)_+} \nonumber \\
     \gamma_{{\bf F}}^{q \rightarrow q+g}(z, \theta_{12})&=& 2\frac{\alpha_s}{\pi}\frac{1}{\theta_{12}}\delta(1-z)
 \end{eqnarray}
 We therefore see a nontrivial mixing between the single medium interaction collinear soft functions. This is to be expected since the higher order corrections to ${\bf F}_{q,1}$ will lead to interference effects between multiple collinear-soft emissions inside the jet giving the same structure as the soft limit of two subjet single medium interaction function ${\bf F}^{a\bar a ii'kk'}_{2,1}$. 
 This shows the power of the EFT framework which allows us capture higher order results without explicit calculations.
Therefore, we now have the LL RG evolution for all the functions computed in this paper except for RG evolution in virtuality $\mu$ for the function ${\bf  F}_{2,1}^{a\bar a ii'kk'}$. This requires a two loop calculation of atleast some of the other functions, which is beyond the scope of this paper but will be very interesting as a follow up. 


\section{Discussion}
\label{sec:Disc}
In this paper we have presented a systematic way to incorporate the phenomenon of color decoherence in jet quenching in an EFT framework. For jet production at threshold, we show that the cross section can be written as a series of multi sub-jet operators that source collinear soft radiation through both vacuum and medium induced interactions. The vacuum evolution of multi sub-jet operators captures the non- global logarithms while the medium induced evolution encodes both the LPM and color decoherence phenomenon. We provide explicit operator definitions for these sub-jet operators within the SCET formalism and compute the one loop matching co-efficients for the single and two sub-jet operators, also deducing their leading order Renormalization Group evolution. We further show that redefining the operators to remove collinear overlap ensures IR finiteness of Wilson co-efficients. We also present the one loop results for the vacuum and single interaction medium evolution for the single and two sub-jet collinear soft functions. Either through explicit computation or by using consistency of factorization, we also present the Renormalization group evolution for most of these functions.

Using these calculations allows us to see how the scale $\theta_c$ emerges in the EFT framework. Further, based on power counting, we conclude that for a jet of radius R, both LPM effect and color decoherence are controlled by a single power counting parameter $\lambda_m = \sqrt{\hat q L}LR$. We conclude that for inclusive jet production, contribution of all the multi sub-jet operators appears at the same order in power counting. So far, these effects have only been studied in simpler settings and this work fills in the crucial gap for a framework that allows for complete calculations for jet substructure observables in Heavy Ion collisions incorporating all known interference driven phenomenon. This can also serve as a template for constructing more realistic parton showers in the medium. 
There are still important steps that need to be taken to get to a phenomenological study using this framework and we detail some of the natural follow up questions that need to be addressed. 

One obvious extension is a two loop computation of the factorized functions, which will allow us to access the RG equations for the two sub-jet collinear soft function.
While we already have the RG equations at Leading Log, a two loop calculation will allow us to solve the RG equations and resum logarithms at NLL which will considerably improve the accuracy of the calculation. 

We note that the main proposal of this work, the factorization formula in Eq.~\ref{eq:factI2}, is valid in both the dense and dilute regimes. There are three important scales in the problem: the mean free path of the jet partons $\ell_{\rm mfp}$, which is related to $\hat q$, the formation time of jet partons $t_F$, and the medium length $L$. For now, the explicit calculations using this framework are restricted to the regime $\ell_{\rm mfp} \gg t_F \sim L$, which is one of the possible hierarchies in the dilute regime. Here, a single interaction is sufficient, and we recover the LPM term  as well as  $\theta_c$, associated with it in this paper. The other important scenario is the dense regime when $L \gg t_F \gg  \ell_{\rm mfp} $, where the parton can exchange several Glauber gluons coherently over its formation time.  At leading order, this leads to the BDMPS-Z result, and radiative corrections lead to a non-linear evolution~\cite{Iancu:2014kga,Blaizot:2014bha}. So far all results in literature in the dense regime primarily use the BDMPS-Z result. However, as we have seen in this work, incorporating color decoherence is just as important if accurate predictions are to be made. Hence a pressing question is to compute color decoherence effects in a dense medium using an EFT framework. 

Another aspect that needs to be considered is the separation of perturbative physics from non-perturbative. In our case, the matching co-efficients ${\cal C}_{i\rightarrow m}$ are perturbative while the same is not necessarily true for the sub-jet function ${\cal S}_{n,k}$ which describes physics at and below the scale $Q_{\text{med}}$. In a dilute medium ($Q_{\med} \sim m_D$), at currently accessible QGP temperatures, we expect this to be non-perturbative while in a dense medium it may be a perturbative scale. Therefore, understanding this separation of scales through an operator based formalism in the dense regime becomes crucial, which is another reason for an EFT approach.

\section*{Acknowledgements}
V.V. is by startup funds from the University of South Dakota and by the U.S. Department of Energy, EPSCoR program under contract No. DE-SC0025545.

\appendix

\section{Operator definitions in SCET}
\label{app:Ocs}

While the operators and Hamiltonians used in the main text can be found in Ref.~\cite{Rothstein:2016bsq}, we collect them here for convenience. All of the SCET operators are defined in terms of gauge invariant building blocks
\small
\begin{align}
 \chi_{n} &= W_{n}^{\dagger}\xi_{n}, 
  \   \   \   \    
  W_{n} = \text{FT} \  {\bf P} \exp \Big\{ig\int_{-\infty}^0 {\rm d}s \, \bar{n}\cdot A_{n}(x+\bar{n}s)\Big\}  ,
  \nonumber\\
 \chi_{\rm s} &= S_{n}^{\dagger}\xi_{\rm s},  
   \   \   \   \    
   S_{n} = \text{FT} \ {\bf P} \exp \Big\{ig\int_{-\infty}^0 {\rm d}s\, n\cdot A_{\rm s}(x+s n)\Big\}
   , \nonumber\\
 & \mathcal{B}_{n \perp}^{\mu }\equiv \mathcal{B}_{n \perp}^{\mu a}t^a  = \frac{1}{g}\Big[W_{n}^{\dagger}iD_{n \perp}^{\mu}W_{ n}\Big], \nonumber \\
 \  \  
 &\mathcal{B}_{\rm s\perp}^{\mu }\equiv \mathcal{B}_{\rm s\perp}^{\mu a}t^a  = \frac{1}{g}\Big[S_{n}^{\dagger}iD_{\rm s \perp}^{\mu} S_{n}\Big] 
 .
\end{align}
\normalsize
Here, FT denotes the Fourier transform. These operators encode bare quarks and gluons dressed by Wilson lines.

Analogous to the quark jet function introduced for the hard, hc factorization, we can also define the gluon jet function as
\begin{align}
&\!J_g(z,\omega_J,\mu)\!=\!\frac{\omega}{4C_FN_c}\int\!\! {\rm d}s\, e^{is\omega}\tr\left[\mathcal{B}_{n\perp\mu}(0)\rho_M\mathcal{B}_{n\perp}^{\mu}(s \bar n)\mathcal{M} \right]. 
\label{eq:gjet}
\end{align}
Further, the subjet function $\mathcal{S}_1$ in the hc, collinear-soft stage of the factorization is evaluated with the Hamiltonian 
\small
\begin{equation}
\label{eq:H2}
\int\! {\rm d}t\, H(t)\!=\!\int {\rm d}t \left(H_{\rm cs}(t)+H_{\rm s}(t)+ H_{\rm cs\text{-}s}(t)\right) \,.
\end{equation}
 Here, $H_{\text{cs}}$ is the standard collinear SCET Hamiltonian, and $H_{\text{s}}$ describes the dynamics of the soft field, which is the full QCD Hamiltonian. Moreover, $H_{\text{cs-s}}$ describes the forward scattering of the collinear-soft gluon off a soft medium parton. 

The Hamiltonian interaction density of the collinear-soft mode with the soft field is given by 
\begin{align}
\label{EFTOp}
\mathcal{H}_{\rm cs\text{-}s}
 &= C_G\frac{i}{2}f^{abc}\mathcal{B}_{n \perp\mu}^b\frac{\bar n}{2}\cdot(\mathcal{P}+\mathcal{P}^{\dagger})\mathcal{B}_{n \perp}^{c\mu}\frac{1}{\mathcal{P}_{\perp}^2}\mathcal{O}_{\rm s}^a
 \,. 
\end{align}
Here, $C_G(\mu)=8\pi\alpha_s(\mu)$ and $\mathcal{O}_{ \rm s}^a =\sum_{j \in \{q,\bar q, g\}}\mathcal{O}_{\rm s}^{j,a}$ with
\small
\begin{align}
\mathcal{O}_{\rm s}^{q, a} &= \bar{\chi}_{\rm s}t^a\frac{\slashed{n}}{2}\chi_{\rm s}, 
 \   \   \    
\mathcal{O}_{\rm s}^{g, a}=  \frac{i}{2}f^{abc}\mathcal{B}_{\rm s \perp\mu}^b\frac{n}{2}\cdot(\mathcal{P}+\mathcal{P}^{\dagger})\mathcal{B}_{\rm s \perp}^{c\mu}. 
 \label{eq:Opbarn}
 \end{align}
The exponent of the operator $U_{\bf M}$ encodes medium-induced radiation to all orders in $\alpha_s$ given by operator $\mathbf{O}_{\text{cs-s}}$, which is defined as 
\begin{align}\label{eq:Oc-s}
\mathbf{O}_{\text{cs-s}}(sn) = \int {\rm d}^2\bfq \frac{1}{\bfq^2}\Big[O^{ab}_{\text{cs}}\frac{1}{\mathcal{P}_{\perp}^2}\mathcal{O}_{\text{s}}^b\Big](sn,\bfq) t^a\,,
\end{align} 
where $O_{\text{cs}}$ and $O_{\text{s}}$ are gauge invariant operators built out of collinear-soft and soft fields, respectively. The operator in Eq.~(\ref{eq:Oc-s}) is given by
\begin{align}
&O_{\rm cs}^{ab} = 8\pi \alpha_s\Bigg[\mathcal{P}_{\perp}^{\mu}S_n^TW_{n}\mathcal{P}_{\perp \mu} -\mathcal{P}_{\mu}^{\perp}g \mathcal{\tilde B}_{\rm s\perp}^{n\mu}S_n^TW_n -\nonumber \\
& S_n^TW_ng \mathcal{\tilde B}^{n \mu}_{\perp}\mathcal{P}_{\mu}^{\perp}- g \mathcal{\tilde B}_{\rm s \perp}^{n\mu}S_n^TW_ng\mathcal{\tilde B}^{n}_{\perp \mu}-\frac{n_{\mu}\bar n_{\nu}}{2}S_n^T ig \tilde G^{\mu\nu}W_n\Bigg]^{ab},
\label{eq:LPT}
\end{align}
\normalsize
where $\tilde G^{\mu\nu}$ is the gluon field strength tensor.

\bibliography{SemiJet.bib}
\end{document}